\documentclass[12pt]{article}

\usepackage{amsmath}
\usepackage{amsfonts}
\usepackage{amssymb}
\usepackage{natbib}
\usepackage{graphicx}
\usepackage{pdfpages}
\usepackage[onehalfspacing]{setspace}
\usepackage[margin=1.5in]{geometry}
\usepackage{subcaption}
\usepackage{setspace}
\usepackage{placeins}
\usepackage{url}
\usepackage{xcolor}
\usepackage{mathtools}
\usepackage{titlesec}

\titlelabel{\thetitle.\quad}

\newcommand{\blind}{1}

\begin{document}

	\def\spacingset#1{\renewcommand{\baselinestretch}%
		{#1}\small\normalsize} \spacingset{1}
	
	\pagenumbering{arabic}

	
	\if1\blind
	{
		\title{\bf Exploring Dependence Structures in the International Arms Trade Network}
		\author{Michael Lebacher\thanks{Department of Statistics, Ludwig-Maximilians-Universit\"at M\"unchen, 80539  Munich,  Germany, michael.lebacher@stat.uni-muenchen.de}\hspace{.2cm} 
			and 
			G\"oran Kauermann\thanks{Department of Statistics, Ludwig-Maximilians-Universit\"at M\"unchen, 80539  Munich,  Germany, goeran.kauermann@stat.uni-muenchen.de}\hspace{.2cm}\thanks{The authors gratefully acknowledge funding provided by the german research foundation (DFG) for the project \textit{International Trade of Arms: A Network Approach}.}}
		\maketitle
	} \fi
	
	\if0\blind
	{
		\bigskip
		\bigskip
		\bigskip
		\begin{center}
			{\LARGE\bf Exploring Dependence Structures in the International Arms Trade Network}
		\end{center}
		\medskip
	} \fi
	
	\bigskip
	\begin{abstract}
		\noindent In the paper we analyse dependence structures among international trade flows of major conventional weapons from 1952 to 2016. We employ a Network Disturbance Model commonly used in inferential network analysis and spatial econometrics. The dependence structure is represented by pre-defined weight matrices that allow for relating the arms trade flows from the network of international arms exchange. Several different weight matrices are compared by means of the AIC in order to select the best dependence structure. It turns out that the dependence structure among the arms trade flows is rather complex and can be represented by a specification that, simply speaking, relates each arms trade flow to all exports and imports of the sending and the receiving state. By controlling for explanatory variables we are able to show the influence of political and economic variables on the volume traded. 
	\end{abstract}
	
	\noindent%
	\vfill
	
	\newpage
	\spacingset{1} 

\section{Introduction}
In this paper we investigate international trade flows of Major Conventional Weapons (MCW) using data provided by the Stockholm International Peace Research Institute (SIPRI). MCW include armoured vehicles, aircrafts, naval vessels etc.\ and SIPRI has collected all international arms transfers from 1950 to 2016 in a comprehensive database.  The database includes information on the sending and the receiving country as well as the volume of the trades in a certain year. The volume is measured in so called TIV, shorthand for trend-indicator value(s), and represents the value of military resources that are exported. The data can be regarded as a year wise sequence of weighted networks where the countries are the nodes and the arms trade flows among them are the valued edges.

Several scholars have started to investigate arms trade using a network framework, but the available studies were restricted to binary relations, i.e.\ trade or no trade, see for example \citet{willardson2013}, \citet{akerman2014}, \citet{Kinne2016} and \citet{thurner2017}. The central workhorse in inferential binary network analysis is thereby the Exponential Random Graph Model (ERGM), as introduced by \citet{holland1981} (see also \citealp{frank1986}, \citealp{kolaczyk2009} and \citealp{lusher2012}). Recent proposals for modelling valued networks within the ERGM class are provided by \citet{krivitsky2012} or the Generalized Exponential Random Graph Model (GERGM) by \citet{desmarais2012}. The first model allows for discrete valued counts while in the GERGM the weights on the edges are transformed into interval data between zero and one. Both approaches are not suitable for modelling of arms trade data because the TIV is continuous and zero inflated (i.e.\ TIV$=0$ stands for no trade). Nonetheless, the binary analysis, i.e.\ trade or no trade, is central, since trading of arms implies governmental agreement and contracts and therefore represents direct and indirect trust relations, regardless of the amount of trading (see e.g.\ \citealp{jackson2010}).

In this paper, however, we focus on the amount of trading (if there is trading). Of particular interest is thereby the dependence structure of trading amounts, that is how is the trading amount of a country (to a country) influenced by other trading amounts. In order to do so, we change the viewpoint of the network and go from a representation with the nodes as countries and the edges being arms trade flows to a network where the nodes are the trade flows between two countries and the edges represent the dependencies among them. This yields a model that at least conceptually resembles models from spatial econometrics  (see e.g.\ \citealp{lesage2009} and \citealp{kauermann2012}). This model class was developed for capturing spatial effects and spillovers from nearby sites and can be transformed to model the dependence of similar arms trade flows. Models from spatial econometrics even have their own name in the social network methodology and are commonly called {\sl Network Autocorrelation Models}  (see for example \citealp{leenders2002} and \citealp{hays2010}). 

The paper is organized as follows. In Section \ref{descrip}, we give a short description of the data, then in Section \ref{model} we introduce our model and give different specifications and explanations for possible dependence structures. In Section \ref{results} we present the results of our analysis. Section \ref{conc} concludes the paper.

\begin{figure}[t!]
	
	\centering
	\includegraphics[trim={0.2cm 0.1cm 0cm 0.5cm},clip,scale=0.8]{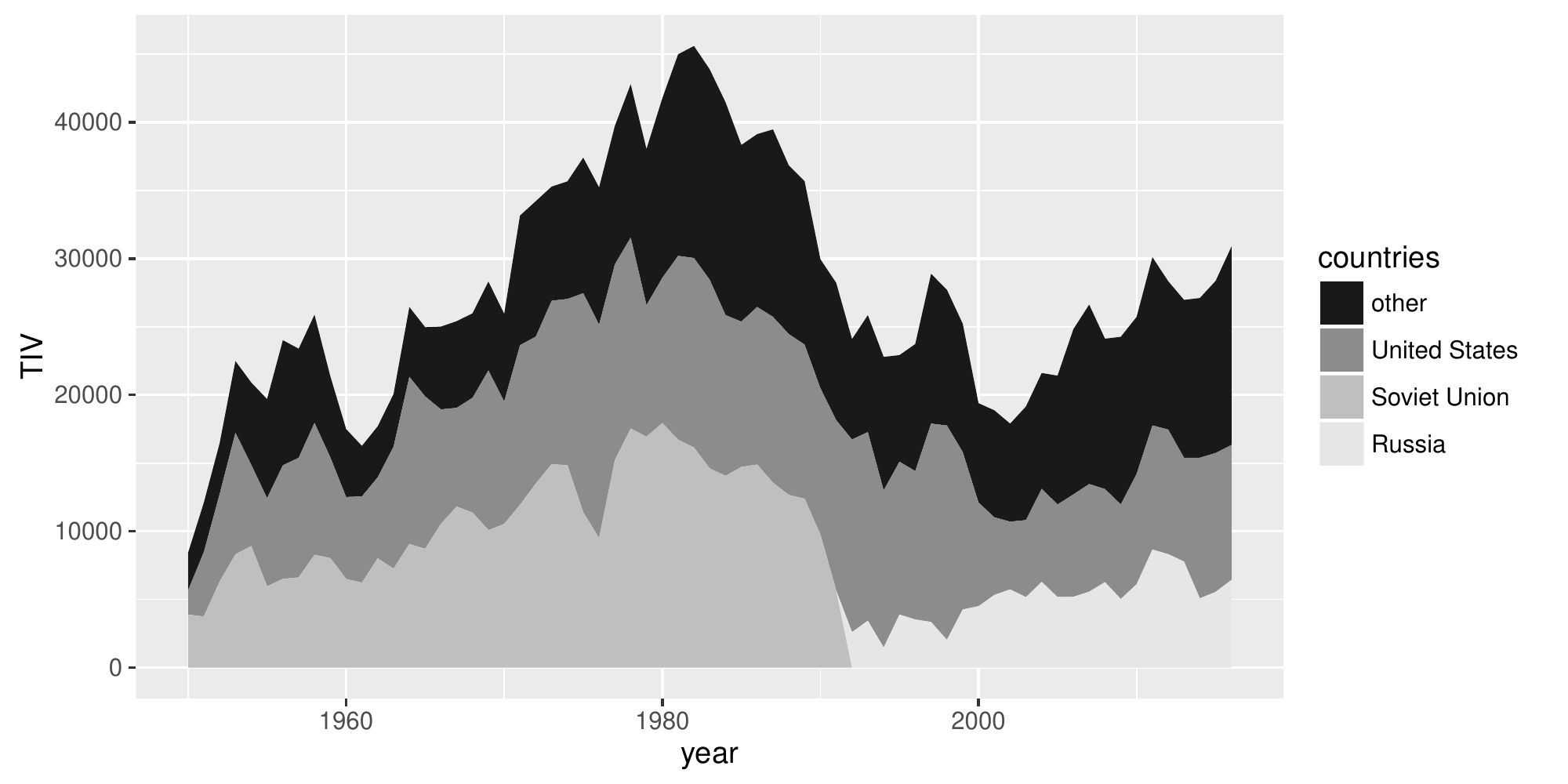}
	\caption{Development of exported TIV for 1950-2016 for the United States, Soviet Union, Russia and all other Countries.}
	\label{stacked}
\end{figure}
\section{Data description}\label{descrip}
\noindent The Stockholm International Peace Research Institute (SIPRI) is a unique data source on the international transfers of major conventional weapons. It covers more than 200 countries for the time period from 1950-2016 and includes information on the sending country, the receiving country as well as the type of MCW that is traded. Table \ref{coverage} in Annex \ref{descrannex} gives an overview of weapon systems included. SIPRI has developed a measure that represents arms trade flows as so called trend indicator values (TIV), based on production costs. The advantage of this measure is consistency over time and comparability of different arms systems. For detailed explanation of the data and methodology see \citet{siprimeth2017} or \citet{holtom2012}.  The database can be accessed free of charge online at \citet{sipridata2017}. 
\begin{figure}
	
	\centering
	\includegraphics[trim={0cm 0.45cm 0cm 0.7cm},clip,scale=0.4]{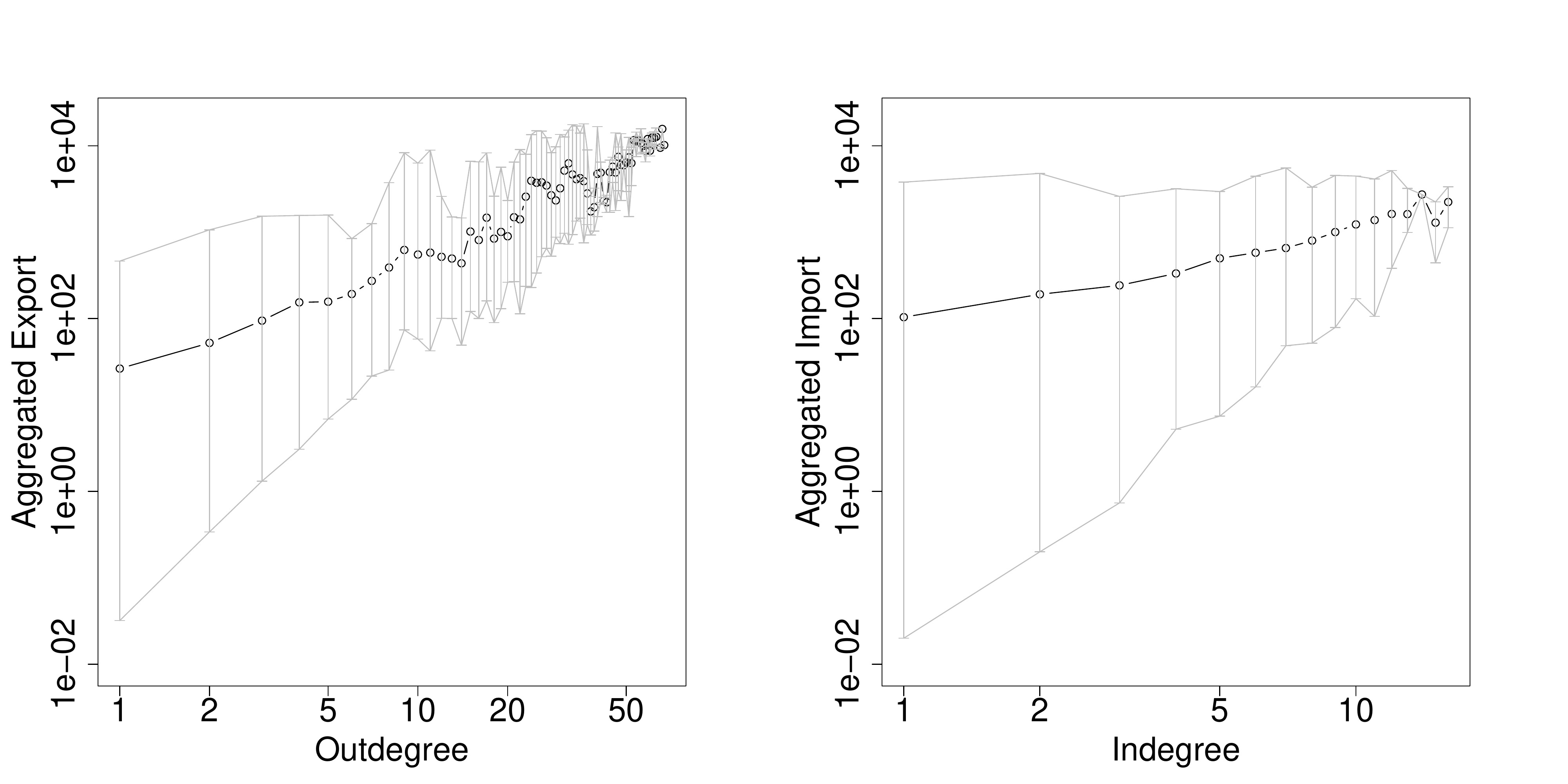}
	\caption{Log-Log Plot of TIV Export vs. Outdegree (left) and TIV Import vs. Indegree (right). Whiskers indicate minimum and maximum Values.}
	\label{degr_dist}
\end{figure}

In Figure \ref{stacked} the development of the aggregated TIV for the time period from 1950 to 2016 is shown for the two most important exporters United States and the Soviet Union (Russia since 1992).
In Figure \ref{degr_dist} a log-log plot of the outdegree and the aggregated export volume on the left and the indegree and the aggregated import volume on the right is given. The solid line in the middle gives the mean of exports or imports that is associated with the respective in- or outdegree, the whiskers indicate the maximum and the minimum value observed. In terms of the slope and range it can be seen that the connection between aggregated export volume and the outdegree is somewhat stronger than the connection between import volume and indegree. Nevertheless, both volumes increase on average in the degree. 
\FloatBarrier
\section{Model}\label{model}
\subsection{Definitions and Basic Formulation}
In $T$ time periods, we observe $n_t$ flows $Y_t =\{Y_{t,ij}:Y_{t,ij}>0, i,j=1,....,N_t\}$ where $Y_{t,ij}$ is the valued arms trade flow from country $i$ to country $j$ among the $N_t$ countries that exist in $t=1,...,T$. In the following, we suppress the time index $t$ for ease of the notation. Let $V=\{(i,j):Y_{ij}>0\}$ be the (directed) index set for all country pairs $(i,j)$ where valued trade flows exists so that $Y=\{Y_v:v \in V\}$. Let $Y$ be of dimension $n=|V|$ and numerate the elements in $V$ so that $V=\{v_1,...,v_n\}$.

As a tool to analyse the dependence structure amongst $Y$ we are using a graph $G$ consisting of nodes and edges, where we take $V$ as node set and define $E$ as symmetric edge set:  $G=(V,E)$.  We define the neighbourhood in the model through 
\begin{equation*}
\mathcal{N}(v_a)\coloneqq\{ v_b: \{v_a,v_b\} \in E   \}\text{, }a=1,...,n \text{.}
\end{equation*}
%
The information given in the neighbourhood can be summarized in a $(n\times n)$ matrix $\mathbf{W}$. Typically the rows of the matrix are normalized such that they sum up to one by defining $
W_{ab}=|\mathcal{N}(v_a)|^{-1}$ if $v_b \in \mathcal{N}(v_a)$ and zero otherwise. This is also the definition we are using. As we are interested in the dependencies among arms trade flows we will henceforth attach the name {\sl tradecorrelation} to the concept of related (neighbouring) flows. The intention of the subsequent analysis is to find a reasonable weight matrix  $\mathbf{W}$, which defines an appropriate dependence structure.

\subsection{Regression Model}\label{models}
Arms trade flows are non-negative, leading us to consider a multiplicative model of the form
\begin{equation}
\label{eq:mod1}
\begin{split}
\log(Y)&=\mathbf{x}^{\sf T}\boldsymbol{\beta}+u \\
(u_1,...,u_n)^{\sf T}=\mathbf{u}&\sim N_n(\mathbf{0}, \mathbf{\Sigma}).
\end{split}
\end{equation}
Here $\mathbf{x}$ represents a vector of some covariates, $\boldsymbol{\beta}$ gives the parameter vector and $u$ is the error term. The residuals $\mathbf{u}$ are assumed to be correlated and $\mathbf{\Sigma}$ provides the variance-covariance matrix encoding the relation of different arms trade flows. Suitable candidates for estimation of (\ref{eq:mod1}) are models from spatial econometrics or spatial statistics, respectively. We will employ the former. Spatial econometric models are common in network analysis, see for instance \citet{leenders2002}, \citet{tnam}, \citet{hays2010}, \citet{metz2017},  \citet{silk2017} and \citet{ba2017}. We follow a Spatial Error Model (SEM, see \citealp{lesage2009}) which is given by defining the error term $\mathbf{u}$ as
\begin{equation}
\begin{split}
\mathbf{u}&=\rho \mathbf{W} \mathbf{u} + \boldsymbol{\epsilon},\\ \boldsymbol{\epsilon}&\sim N_n(\mathbf{0}, \sigma^2\mathbf{I}_n),
\end{split}
\label{SEM}
\end{equation}
where $\mathbf{I}_n$ is  the identity  matrix of dimension $n$ and $\rho$ provides a measure for the dependency among the related error terms $\mathbf{u}$.
The error terms $\boldsymbol{\epsilon}$ are assumed to be unstructured so that equation (\ref{SEM})  can be reformulated to
\begin{equation}
\mathbf{u}=(\mathbf{I}_n-\rho \mathbf{W})^{-1}\boldsymbol{\epsilon}.
\label{error}
\end{equation} 
This implies the covariance matrix $\boldsymbol{\Sigma}= \sigma^2(\mathbf{I}_n-\rho \mathbf{W})^{-1}(\mathbf{I}_n-\rho \mathbf{W}^{\sf T})^{-1}$. Hence, matrix $\mathbf{W}$ expresses the dependence structure and the explicit choice of $\mathbf{W}$ is the central model selection task that will be discussed in Section \ref{weightmat}. The modelling approach is also known as  {\sl Network Disturbance Model} (NDM) in the network literature, delegating the dependencies among trade flows into the error term. Hence, the approach does not aim to model a direct dependence between the trade flows but implies that the exogenous covariates determine the "right" value for an arms trade flow, but one can find clusters of trade flows where the residuals are unusually high or low (see \citealp{leenders2002}  and \citealp{freeman2017} for further discussion). We sketch some details on estimation and software in the Annex \ref{estimation_annex}.
\subsection{Specifications of the Weight Matrix}\label{weightmat}
\begin{figure}[t!]
	
	\centering
	\includegraphics[trim={3.2cm 3.5cm 2.2cm 3.3cm},clip,scale=0.55]{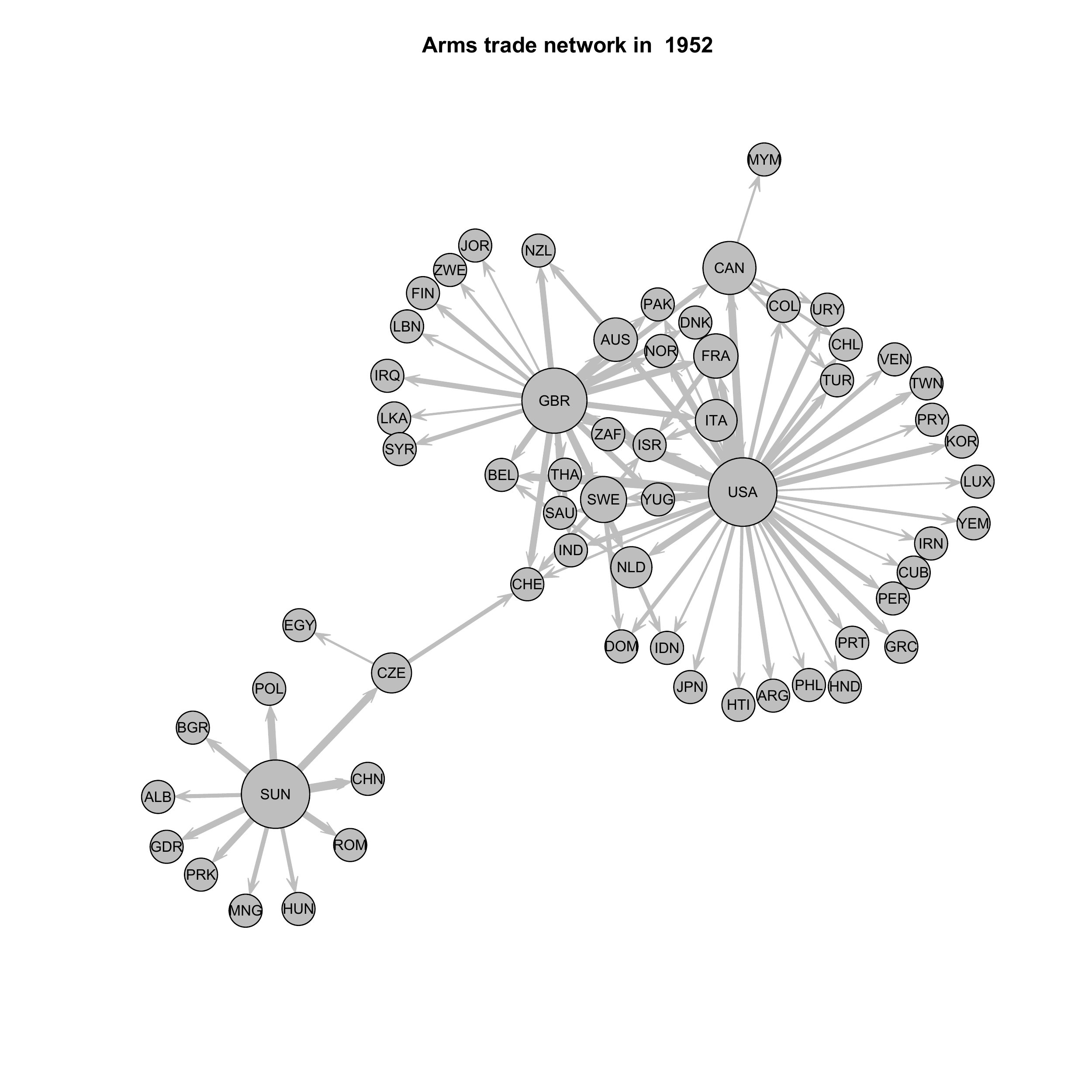}
	\caption{Directed valued Network of Arms Trade in 1952. Edges (Nodes) are scaled proportional to the logarithmic Exports (logarithmic aggregated Exports) measured in TIV.}
	\label{net52}
\end{figure}

We now focus on the central modelling task which is the appropriate specification of matrix $\mathbf{W}$ and hence the definition of the neighbourhood $\mathcal{N}$. We thereby focus on interpretable structures. Given that in the network context the weight matrices are rather abstract, we illustrate our approach with the arms trade network of the year 1952, being the simplest network under study. The network is shown in Figure \ref{net52}. Note that all following illustrations of weight matrices $\mathbf{W}$ have their origin in the network shown in Figure \ref{net52} and are therefore not representative for other years. Still, they allow to visually understand the dependence structure.
\FloatBarrier
\begin{figure}[t!]
	\centering
	
	\includegraphics[trim={0.08cm 0.2cm 0cm 0.05cm},clip,scale=0.51,page=1]{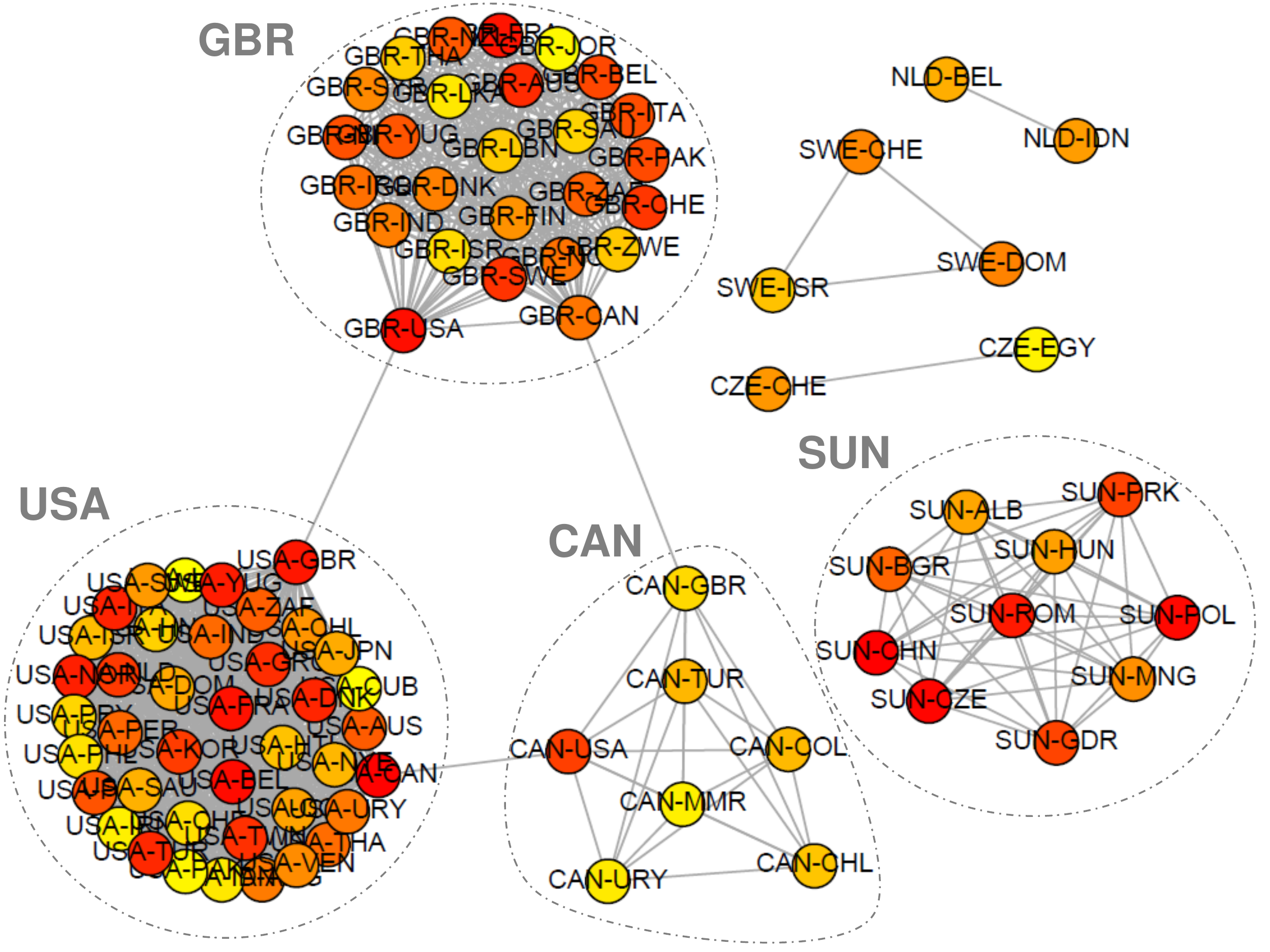}
	
	\caption{Tradecorrelation Structure 1. Colouring according to  logarithmic TIV, ranging from Yellow (low) to Red (high).}
	\label{tc_1}
\end{figure} 
\FloatBarrier     

\noindent {\sl Tradecorrelation 1 (attached to the Sending Country)}

\noindent    The first dependence structure relates an arms trade flow from country $i$ to country $j$ to all other exports going out of country $i$ and a possible reciprocal trade flow $(j,i)$. Notationally this means
\begin{equation*}
\mathcal{N}_1((i,j))=\{(p,q): p=i \vee (p,q)=(j,i), (p,q)\in V \}.
\end{equation*}
The corresponding graph representation for the year 1952 is shown in Figure \ref{tc_1}. The trade flows are grouped within clusters that are related to single exporting countries, most notably, the trade clusters of the big exporters: United States (USA), United Kingdom (GBR), Canada (CAN) and the Soviet Union (SUN). Via the dependence on reciprocal flows, the clusters are connected among each other if there is between-cluster trade. This explains the connection between the cluster of the United States (USA), Great Britain (GBR) and Canada (CAN). The trade activity of the Soviet Union (SUN), however, is disconnected.

\begin{figure}[t!]
	\centering
	
	\includegraphics[trim={0.1cm 3.5cm 0.1cm 1.2cm},clip,scale=0.6,page=2]{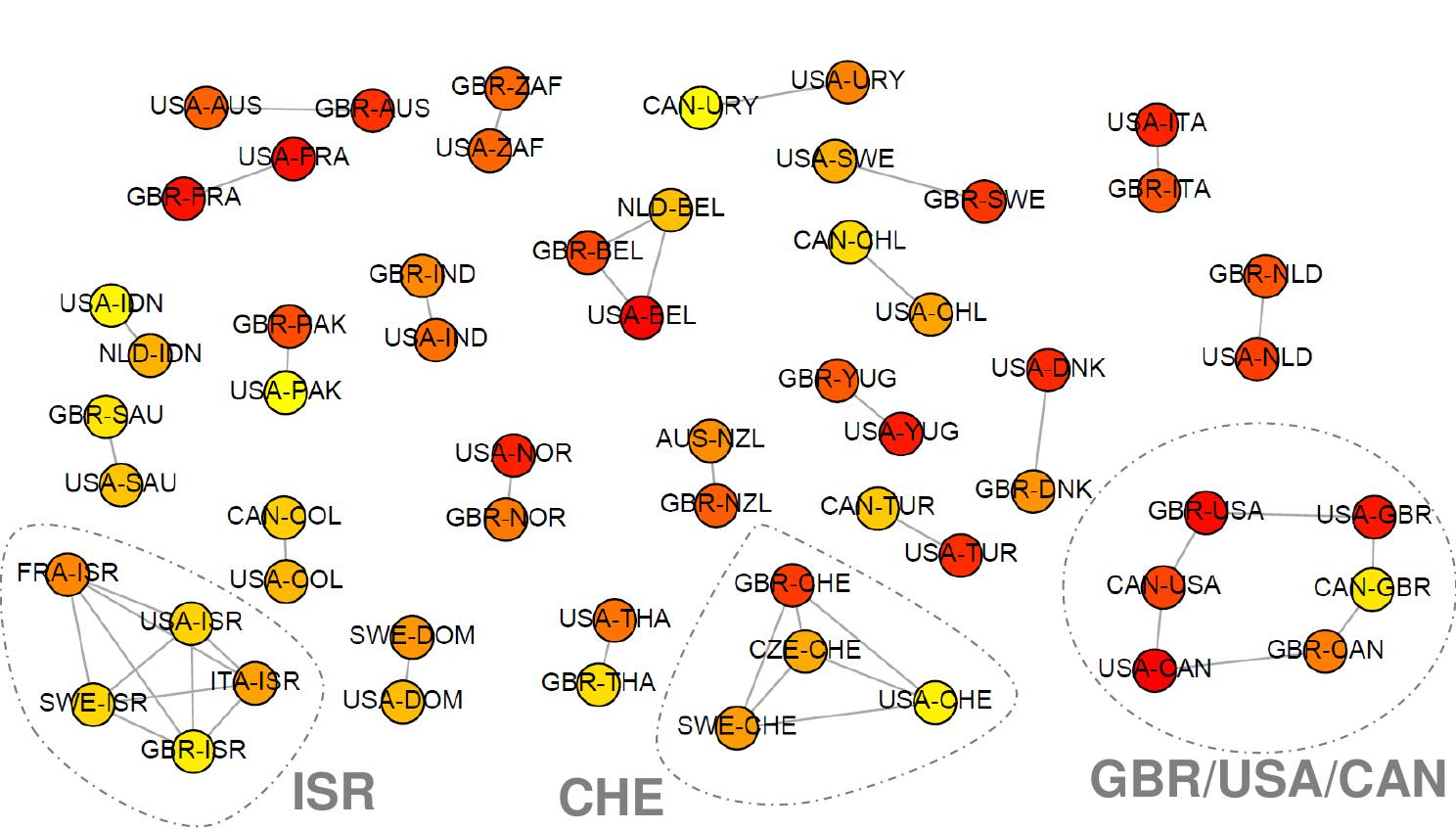}
	
	\caption{Tradecorrelation Structure 2. Colouring according to logarithmic TIV values, ranging from Yellow (low) to Red (high).}
	\label{tc_2}
\end{figure}
\newpage
\noindent{\sl Tradecorrelation 2  (attached to the Receiving Country)}

\noindent The concept from above can be amended such that the tradecorrelation of an arms trade is related to the trades of the importing country and not the exporting one. This means \begin{equation*}
\mathcal{N}_2((i,j))=\{(p,q): (p,q)=(j,i) \vee q=j, (p,q)\in V \}
\end{equation*}
and implies that the arms trade amount from $i$ to $j$ is correlated with other imports of $j$ as well as the potential reciprocal trade flow $(j,i)$. This is visualized in Figure \ref{tc_2}. The resulting graph representation is very different from the previous one. There is no tendency of strong clustering as most importing countries do only have one or two suppliers and no exports at all which excludes connections via reciprocal flows. The greatest clusters consist of arms exports to Israel (ISR) and imports of Switzerland (CHE). Great Britain (GBR), United States (USA) and Canada (CAN) show their strong connections in arms trade by representing their own import-related cluster. Note that this structure disregards an important characteristic of the arms trade network: Receiving countries are often restricted to be dependent on one single exporter. Most important, trade flows of the Soviet Union (SUN) are missing.
\newpage
\begin{figure}[t!]
	\centering
	
	\includegraphics[trim={0.2cm 4.8cm 0.2cm 1.9cm},clip,scale=0.6,page=3]{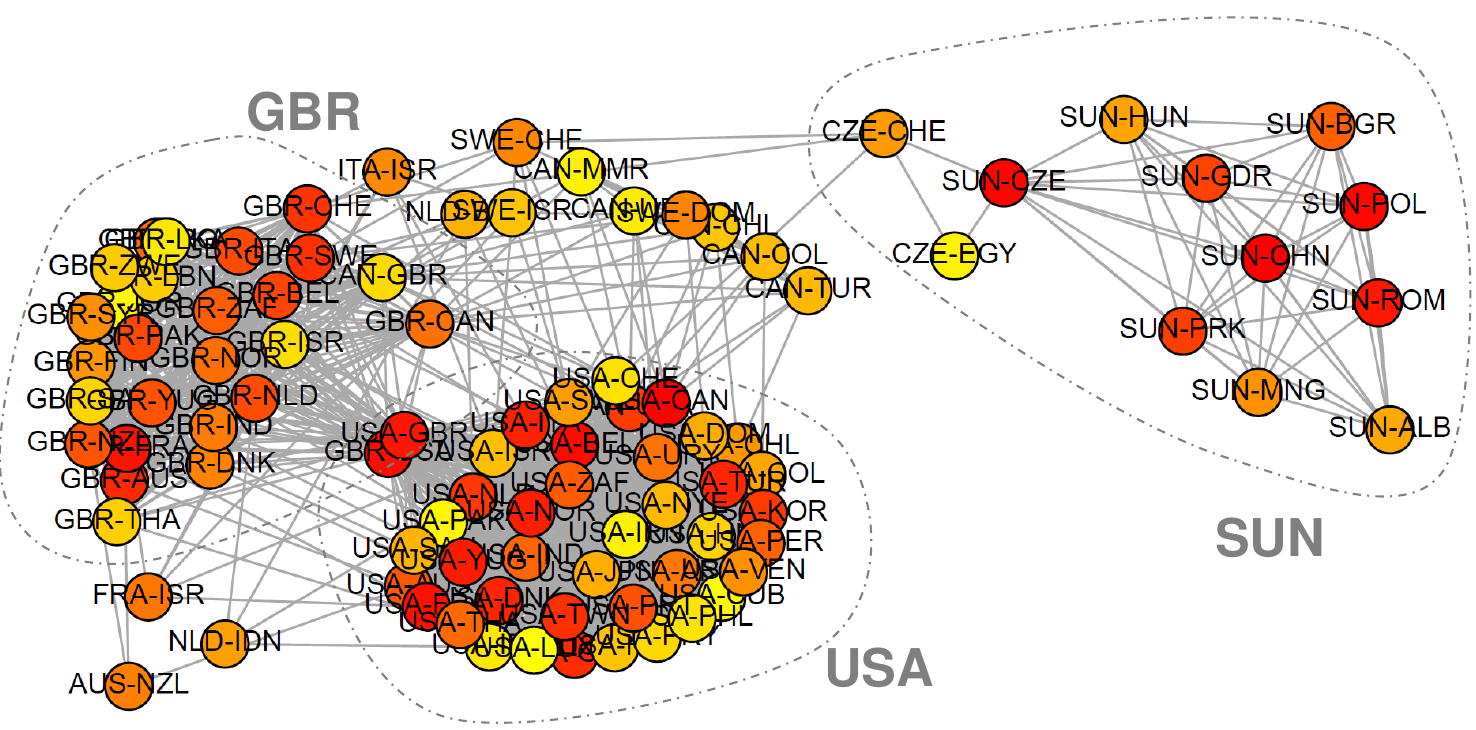}
	
	\caption{Tradecorrelation Structure 3. Colouring according to logarithmic TIV, ranging from Yellow (low) to Red (high).}
	\label{tc_3}
\end{figure}
\noindent {\sl Tradecorrelation 3 (attached to the Trade Activity of Sender and Receiver)}

\noindent        The structure of tradecorrelation 1 and 2 can be combined by the inclusion of the whole trade activity of $i$ and $j$, that is 
\begin{equation*}
\mathcal{N}_3((i,j))=\{(p,q): p\in\{i,j\} \vee q\in\{i,j\}, (p,q)\in V \}.
\end{equation*}
In this case an arms trade from country $i$ to country $j$ is dependent on the whole trade activity of countries $i$ and $j$, regardless whether they are reciprocal or not. This results in a very dense graphical representation, as shown in Figure \ref{tc_3}. Here, the clusters are again very clearly visible but the interconnection within and between clusters is much stronger. Central are the trade flows  United States (USA) - Great Britain (GBR)
and  Great Britain (GBR)    - United States (USA) as they link the two countries with the most trade activity.   The flows related to Switzerland (CHE) serve as a bridge between the clusters around the western producers and the Soviet Union (SUN).  With this specification the graph is connected and in principle each trade flow is allowed to be correlated with an arbitrary other arms trade flow.
\newpage
\begin{figure}[t!]
	\centering
	
	\includegraphics[trim={0.2cm 2.9cm 0.2cm 0.6cm},clip,scale=0.5,page=4]{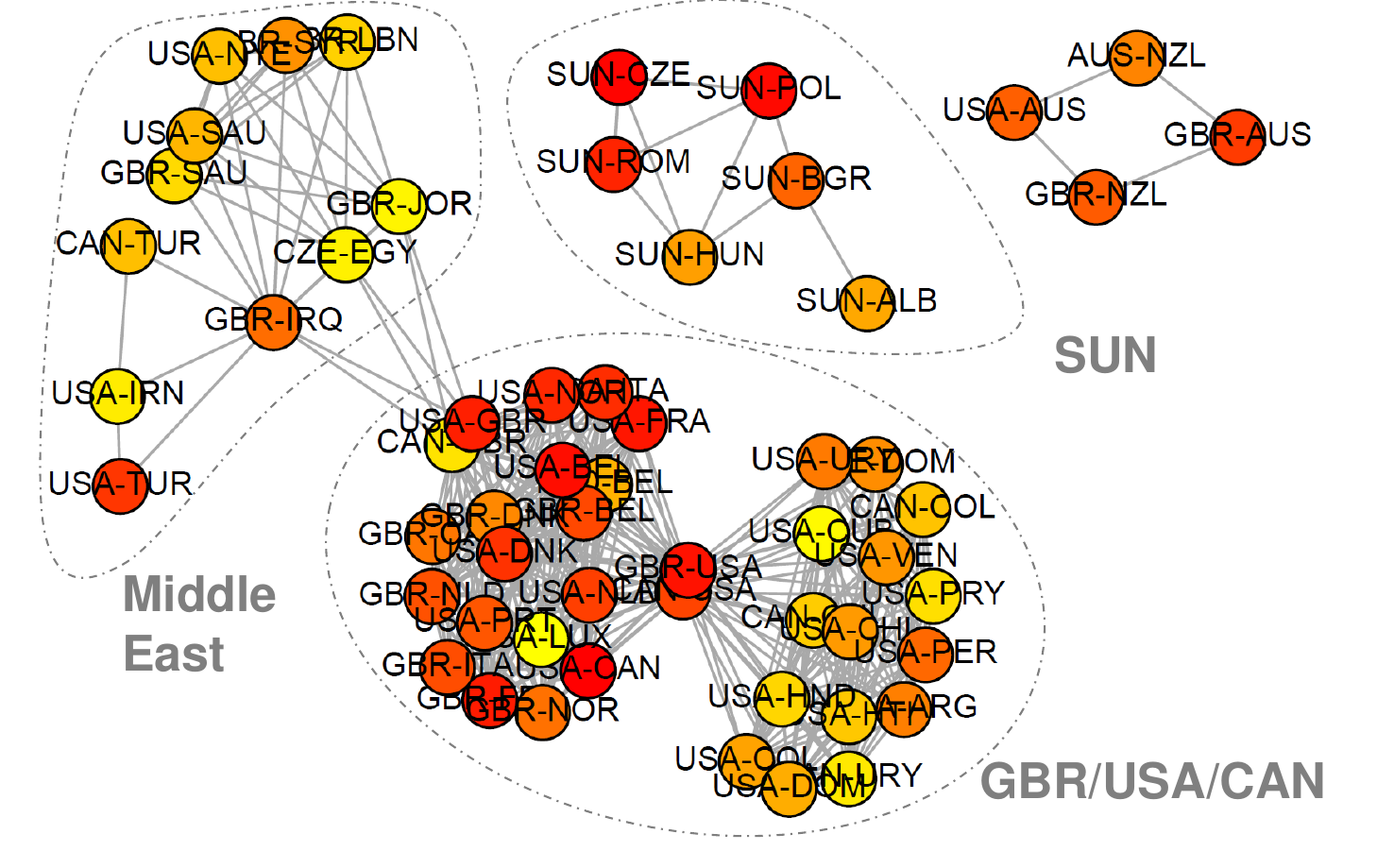}
	
	\caption{Tradecorrelation Structure 4. Colouring according to logarithmic TIV, ranging from Yellow (low) to Red (high).}
	\label{tc_4}
\end{figure}
\noindent{\sl Tradecorrelation 4 (related to Defence Agreements)}

\noindent It can be discussed whether arms exchange is related to military alliances as they might  agree to build up comparable military strength which potentially leads to correlated imports. The \citet{Defagr2016} provides data on formal alliances.
We define a structure that relates the imports of country $j$ to imports of its allies. Let $da(j,i)=da(i,j)$ be an indicator that is one if countries $i$ and $j$ have a formal alliance. Then we define  
\begin{equation*}
\mathcal{N}_{4,import}((i,j))=\{(p,q): da(j,q)=1, (p,q)\in V\}.
\end{equation*} This assumption imposes a structure where importers regard the arms imports of their allies when buying arms. In the corresponding plot in Figure \ref{tc_4} the trade clusters are not any longer attached to the sending countries but are rather clustered among countries that are connected in the alliances network. We see that the United States (USA), United Kingdom (GBR) and Canada (CAN) represent a (bipartite) cluster with their closest allies while the arms trade activity of the Soviet Union (SUN) forms a disconnected cluster. With this structure we additionally see a cluster containing the alliances of Canada (CAN), United Kingdom (GBR) and the United States (USA) together with countries from the Middle East. Naturally it is also possible to relate the exports of sending countries that have an alliance, resulting in
\begin{equation*}
\mathcal{N}_{4,export}((i,j))=\{(p,q): da(i,p)=1, (p,q)\in V\}.
\end{equation*}
\newpage
\begin{figure}[t!]
	\centering

	\includegraphics[trim={0.2cm 0cm 0.2cm 0.15cm},clip,scale=0.52,page=5]{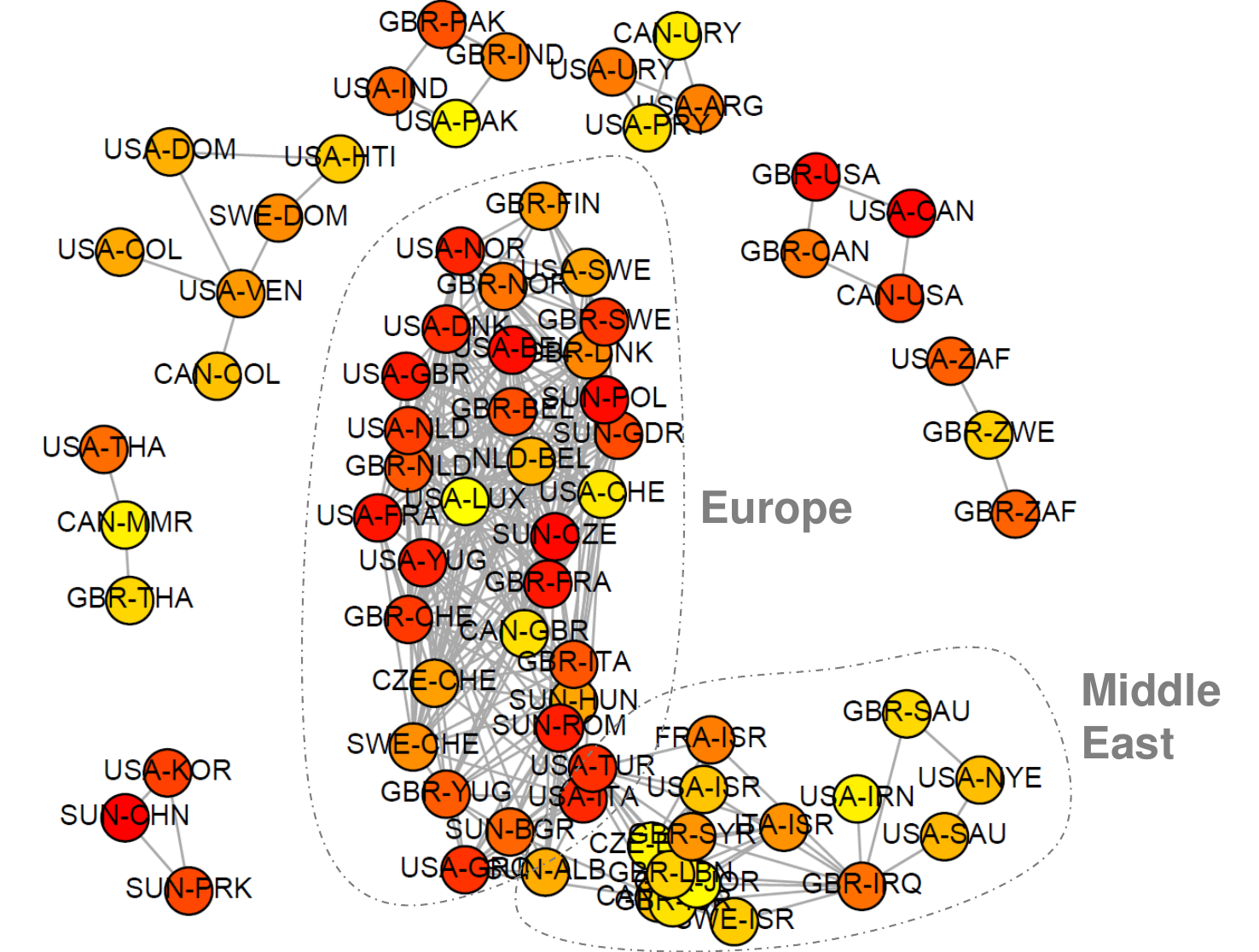}

	\caption{Tradecorrelation Structure 5. Colouring according to logarithmic TIV, ranging from Yellow (low) to Red (high).}
	\label{tc_5}
\end{figure}

\noindent {\sl Tradecorrelation 5 (related to Spatial Distance)}

\noindent Using data provided by \citet{gleditsch2013d} we can incorporate spatial distances in the correlation structure. We assume that arms imports into spatially close areas are correlated and define $d(i,j)=d(j,i)$ as the distance between the two capital cities of $i$ and $j$. With $c$ representing a threshold we define 
\begin{equation*}
\mathcal{N}_{5,import}((i,j))=\{(p,q):j\neq q \wedge d(j,q)<c, (p,q)\in V\}.
\end{equation*}  
Setting $c=1\,100$ km leads to spatially related trade activity clusters as shown in Figure \ref{tc_5}, where in middle a cluster of tradeflows starts with countries from western Europe and relates them over eastern Europe to the Middle East. The choice of $c$ is discussed in the Annex \ref{cutoffrannex}. Note that it is also possible to relate flows of spatially close exporters, that is
\begin{equation*}
\mathcal{N}_{5,export}((i,j))=\{(p,q):i\neq p \wedge d(i,p)<c, (p,q)\in V\}.
\end{equation*}

\FloatBarrier
\subsection{Exogenous Covariates}
We control for the influence from economics and politics by including additional covariates in model (\ref{eq:mod1}). Since there is an average time lag of roughly two years between ordering and delivering of weapons and some covariates are only available from 1950 on, the first network under study is the one of 1952 and covariates are lagged by two years.
\\
{\sl Economic Quantities}: The standard measure for economic size is the {\sl gross domestic product} (GDP). We include this measure in logarithmic form for the sender as well as the receiver. The real GDP data is measured in thousands USD and are taken from \citet{gleditsch2013}. The data are merged from the year 2011 on with recent real GDP data from the World Bank real GDP dataset (see \citealp{GDP2017}). It seems plausible that the {\sl military expenditures} of the receiving countries are relevant. The data is available from \citet{cinc2017} with \citet{singer1972} as the basic reference for the data. As the time series for military expenditure ends in 2012 we have to assume that the values stay constant until 2014.
\\
{\sl Political Quantities}: We use a dummy variable being one if the countries $i$ and $j$ have a {\sl formal alliance} and zero otherwise.  Given the restriction that the data is available only until 2012 (\citet{Defagr2016} for the most recent version of the data and \citet{gibler2008} for the article of record for the data set) we assume that the alliances did not change between 2012 and 2015. In order to control for {\sl regime dissimilarity}, we include the absolute difference of the so called polity score, ranging from the spectrum $-10$ (hereditary monarchy) to $+10$ (consolidated democracy). The data can be downloaded as annual time-series. See \citet{marshall2017} for the data and basic reference.

\FloatBarrier

\section{Results} \label{results}
\subsection{Model Selection}
\FloatBarrier
Our ultimate goal is to select a suitable dependence model from above. As recommended by \citet{leenders2002} for this model class with several candidate weight matrices, we compare the proposed models using the Akaike Information Criteria (AIC) (see for example  \citealp{claeskens2008}). For comparison we also include a simple linear model with unstructured error term (OLS, $\rho=0$). Estimation is done year wise and the AIC is calculated for all estimated models from 1952 to 2016, all in all covering $18\,964$ observations.
\begin{table}[t!] \centering \centering
	\resizebox{\columnwidth}{!}{%
	\begin{tabular}{lp{1.4cm}p{1.4cm}p{1.4cm}p{1.4cm}p{1.4cm}p{1.4cm}p{1.4cm}p{1.4cm}}
		
		\\[-1.8ex]
		\hline \\[-1.8ex] 
		& $\mathcal{N}_1$ &  $\mathcal{N}_2$ &  $\mathcal{N}_3$ &  $\mathcal{N}_{4,import}$& $\mathcal{N}_{4,export}$&  $\mathcal{N}_{5,import}$& $\mathcal{N}_{5,export}$& $\rho=0$ \\ 
		\hline \\[-1.8ex] 
		$\Delta_i$ &$179.1$ & $905.6$ & $0.0$ & $690.7$ & $354.1$ & $816.6$ & $954.6$ & $940.6$ \\ 
		\hline \\
	\end{tabular} 
}•
	\caption{Difference between aggregated AIC Values and the minimum aggregated AIC Value for different Tradecorrelation Structures.} 
	\label{AIC_table} 
\end{table} 
In Table \ref{AIC_table} the results for the  aggregated AIC, rescaled by the lowest one ($\Delta_i=$AIC$_i- $AIC$_{min}$) are presented. AIC$_{min}$ corresponds to tradecorrelation structure 3.
\begin{figure}[t!]
	\centering
	
	\includegraphics[trim={1.2cm 0cm 0cm 0.2cm},clip,scale=0.4]{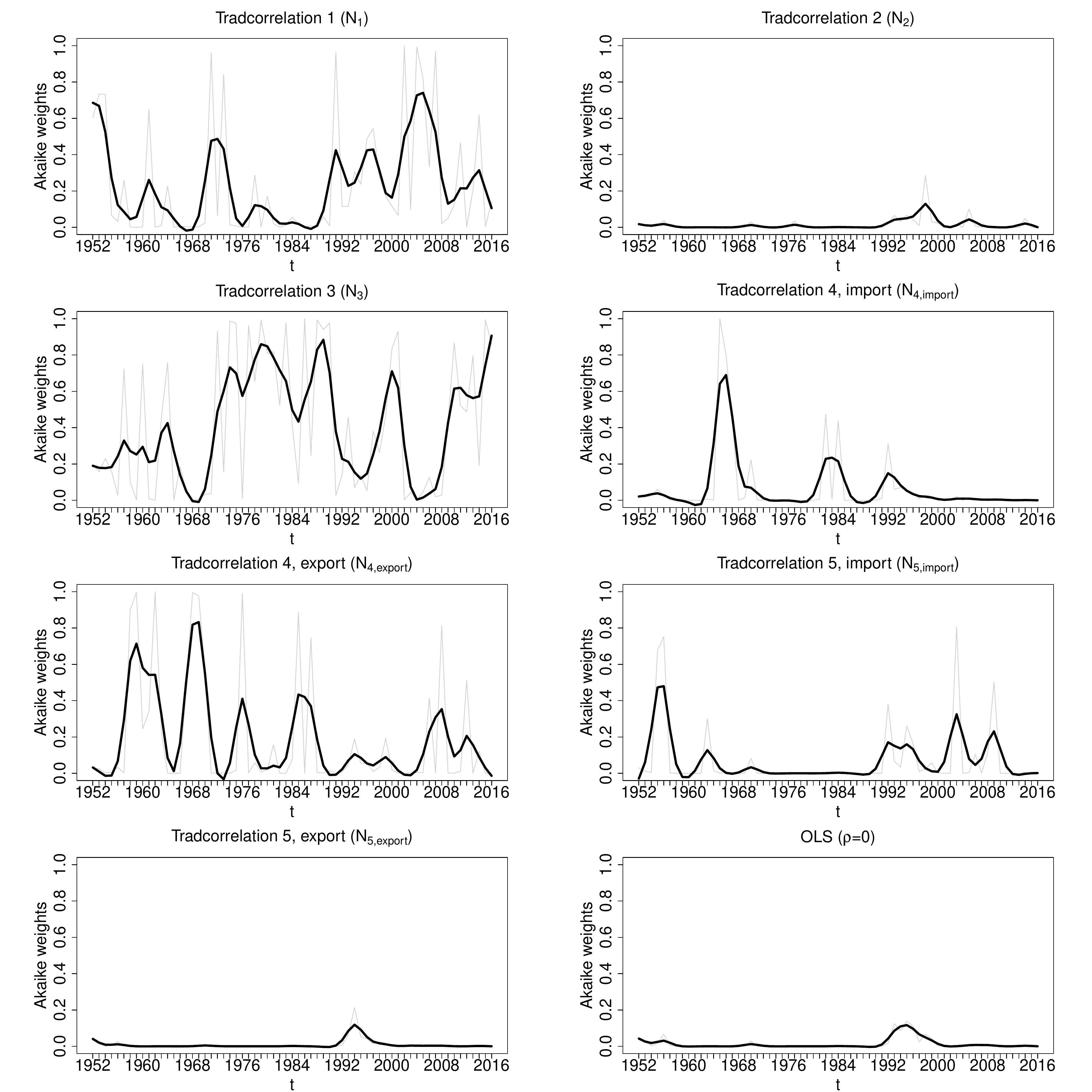}
	
	\caption{Year wise Akaike Weights for different Specifications (Raw Values in Gray, Smoothed in Black).}
	\label{aic_comp}
\end{figure}
In order to investigate the different models more in detail, we also look at year wise AIC values. To do so, we compute Akaike weights, which can be interpreted as the probability of model $i$ being the best (i.e.\ minimizing the Kullback-Leibler Divergence given the data) among the proposed ones. The weights are given by $w_i=\exp(-\Delta_i/2)/(\sum_{r=1}^{R}\exp(-\Delta_r/2))$ (see \citealp{burnham2004} for a further discussion).  The weights are shown in Figure \ref{aic_comp} for all eight specifications.

We see that the specification where $\rho$ is restricted to zero (OLS) is among those models that have a low probability of being the best one. Only in the years 1990-2000 the probability for this model rises slightly as a consequence of the collapse of the Eastern Bloc. Overall, this provides evidence for our initial hypothesis that the trade flows are indeed correlated. 

Specification $\mathcal{N}_{5,export}$ (dependency of exports from spatially close countries), shows almost identical Akaike weights as the OLS specification. A plausible result, given that the capital cities of the big exporters United States and Soviet Union/Russia do not have nearby capital cities of other strong arms exporters and hence the greatest share of trade flows is restricted to be uncorrelated. This stands in contrast to $\mathcal{N}_{5,import}$ (dependency of imports of spatially close countries) that shows up to be among the better models in the beginning of the observational period and after 1989. Hence, this indicates that arms imports into spatially near countries are indeed related to some extent after 1989, which may be interpreted as a spatial clustering of arms imports.

While in tradecorrelation structure 5, the importer related assumption ($\mathcal{N}_{5,import}$) gives the higher probability to be the best one as compared to the exporter related assumption ($\mathcal{N}_{5,export}$), the opposite is the case for tradecorrelation structure 4, that is related to formal alliances. Here the model that assumes dependency of arms trade flows of allied exporting countries ($\mathcal{N}_{4,export}$) performs better than the one that relates those of allied importing countries ($\mathcal{N}_{4,import}$). This leads to the presumption that formal alliances mostly consist of partners with rather similar possibilities to export arms, i.e.\ either high ones, which leads to exports with high residuals or low ones, leading to exports with low residuals. The importers that are connected via formal alliances, seem to be more heterogeneous with respect to the height of their imports. 

For the tradecorrelation structure 2 ($\mathcal{N}_{2}$, relating the imports and a potential reciprocal arms trade flow), we find a similar behaviour as for OLS and $\mathcal{N}_{5,export}$. Leading to the conclusion that there is no strong evidence that the importing arms flows are strongly correlated. 

The two specifications that compete about being the best model are tradecorrelations $\mathcal{N}_{1}$ and $\mathcal{N}_{3}$. They alternate in being the best model in most time points. Mostly if the probability of the model with tradecorrelation $\mathcal{N}_{1}$ decreases, the one with tradecorrelation $\mathcal{N}_{3}$ increases and vice versa. But from a global perspective it is tradecorrelation $\mathcal{N}_{3}$  that provides the best model, which leads to the conclusion that a single arms trade flow is related to the whole trade activity of the sender as well as the receiver.

We infer that the exchange of international arms leads to a rather complex dependence structure. This shows that, similar to the binary analysis, the assumption of independent observations, given the covariates, is a strong simplification of the real process. Given this result we further analyse the model with tradecorrelation structure $\mathcal{N}_{3}$. 

\FloatBarrier

\subsection{Coefficients}
\begin{figure}[t!]
	
	\centering
	\includegraphics[trim={0.25cm 0.2cm 0cm 0.2cm},clip,scale=0.5]{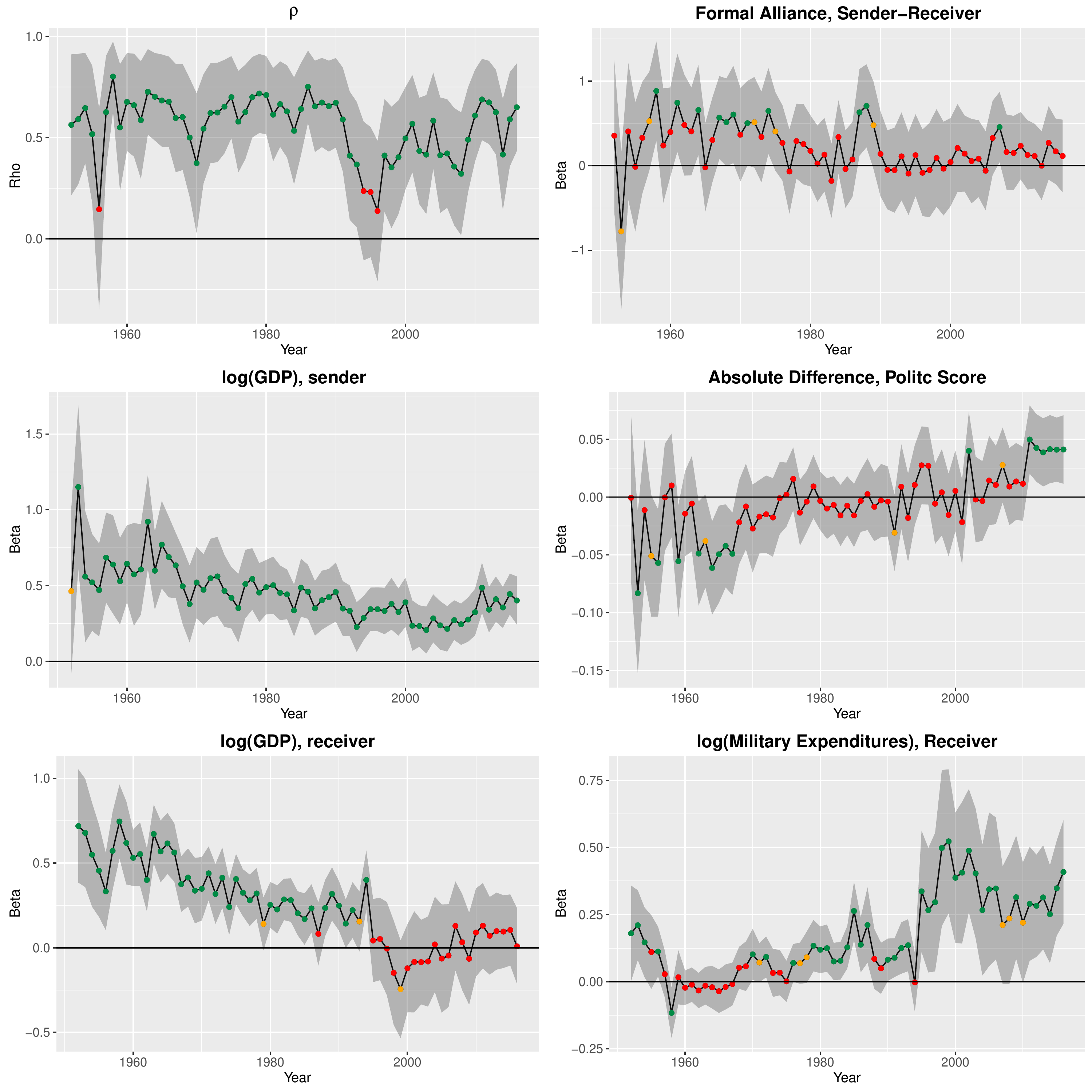}
	\caption{Time Series of Coefficients from the Model with Tradecorrelation Structure 3. $p$-values by colours (green $p<0.05$, orange $p<0.1$, red $p>0.1$). Shaded Regions give $\pm$ 2 standard deviations.}
	\label{coef_fixeff}
\end{figure}

{\sl Tradecorrelation:} The top left panel of Figure  \ref{coef_fixeff} shows the trade correlation coefficient using the structure $\mathcal{N}_{3}$. The coefficient  is consistently positive but exhibits some variation. The strongest downward movement can be found in the years 1990 - 1996, showing how massively the end of the cold war altered the structure of the international trade of arms. From 1996 on, the coefficient increases again. In the most recent years the tradecorrelation fluctuates around $0.6$, close to the values before 1990.
\\
{\sl Economic Quantities:} The coefficient on the logarithmic GDP of the sender is, with the exception of the year 1952, consistently positive and significant, showing a slight tendency to decrease with time. This can be seen from Figure \ref{coef_fixeff}, left panel, middle plot and reflects the fact that mostly wealthy countries have a highly developed arms industry which are able to produce and export expensive military equipment. The slight decrease of the coefficient is probably related to the increase of exporting countries that have smaller economic size. When it comes to the GDP of the receiver (left panel, bottom plot), we see a more pronounced downward trend of the coefficients. It is interesting to observe that from 1994 on, the coefficient becomes permanently insignificant. All in all there is no strong evidence that the TIV of arms imports is strongly influenced by the GDP of the receiving country since the 90s. It is, however influenced by the logarithm of military expenditure of the respective country as can be seen in the bottom plot of the right panel. There the coefficient on the logarithmic military expenditures is almost always positive since 1967 and mostly significant from 1990 on. This indicates a change in the arms trading mechanism. Before the end of the cold war, high arms trade flows went to countries with big economic size and after that, high flows are directed more strongly towards countries with high spending for military equipment.
\\
{\sl Political Quantities:} Before 1990, having an alliance (right panel, top plot) in tendency increased the amount of arms trading but at least since 1990 this effect is almost zero. The coefficient on the absolute differences of the polity score (right panel, middle plot) shows a time related trend. In the beginning (1953-1973) the coefficient is rather less than zero, giving the intuitive result that differences in political regimes reduce the amount of arms traded. From 1976 until 1991 the coefficient can be regarded as zero, which mirrors the fact that at the  height of the cold war the belonging to a political bloc was more important then distances in the polity scores. After the end of the Soviet Union, the coefficient remained close to zero until 2011 where the coefficients becomes significantly positive for six consecutive years, providing the surprising result that differences in the polity score do not lead to a reduction of the traded amount of arms.
\FloatBarrier
\subsection{Residual Analysis}\label{insample}
\begin{figure}[t!]
	\centering
	\includegraphics[trim={0.15cm 1.5cm 0cm 2cm},clip,scale=0.6]{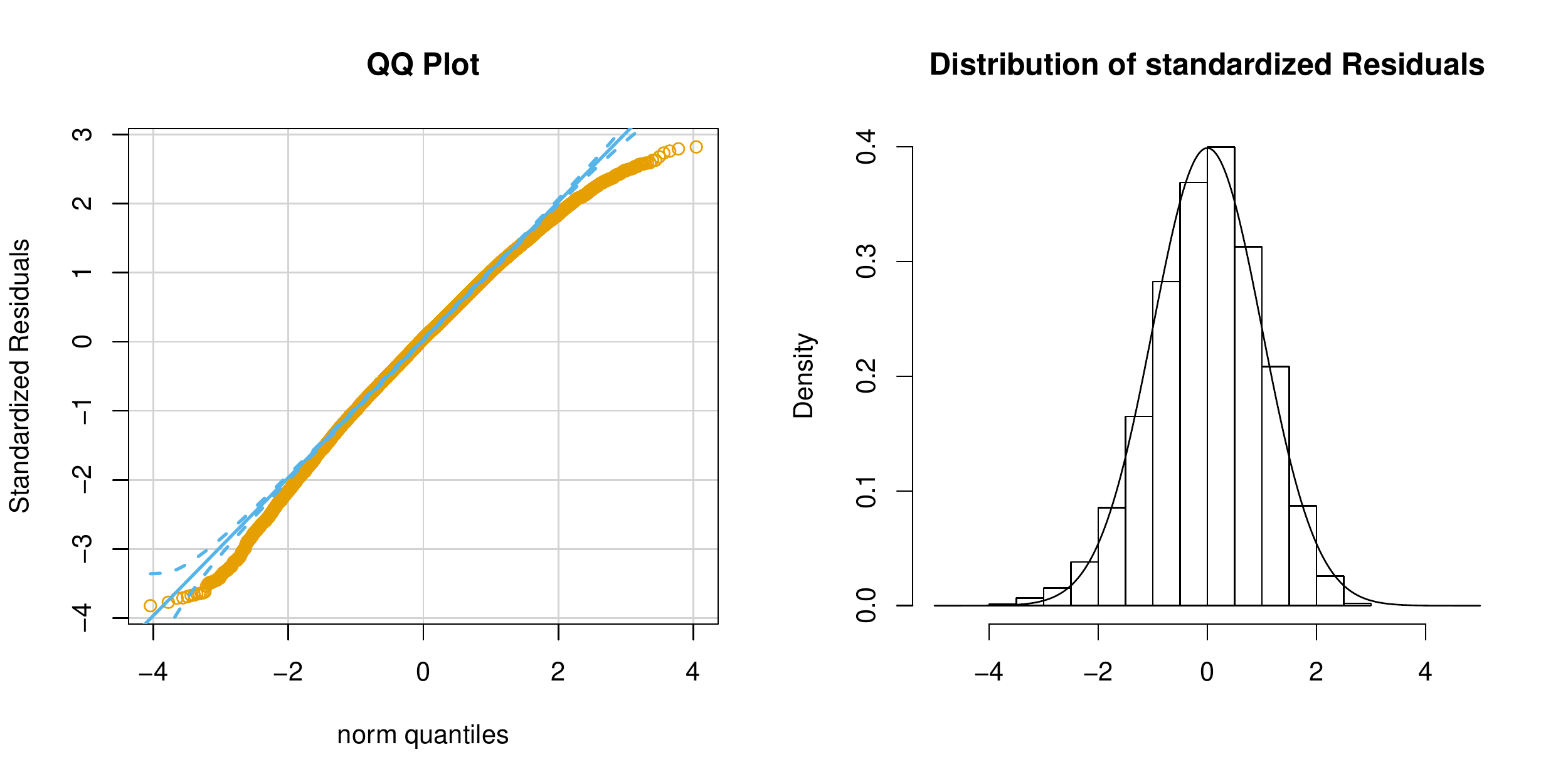}
	\caption{QQ-Plot (left) and Histogram (right) of standardized Residuals.}
	\label{res_diag}
\end{figure}

\noindent We give a graphical summary in Figure \ref{res_diag} for the pooled unstructured estimated residuals $\hat{\boldsymbol{\epsilon}}$. The plot on the left gives a QQ-plot with the quantiles of the standardized residuals on the vertical axes against the quantiles from a standard normal distribution on the horizontal axis. The right panel shows a histogram of the residuals together with a standard normal. Besides some irregularities in the tails, the results look quite acceptable.

\begin{figure}[t!]
	\centering
	\includegraphics[trim={0.15cm 1.3cm 0cm 0.7cm},clip,scale=0.6]{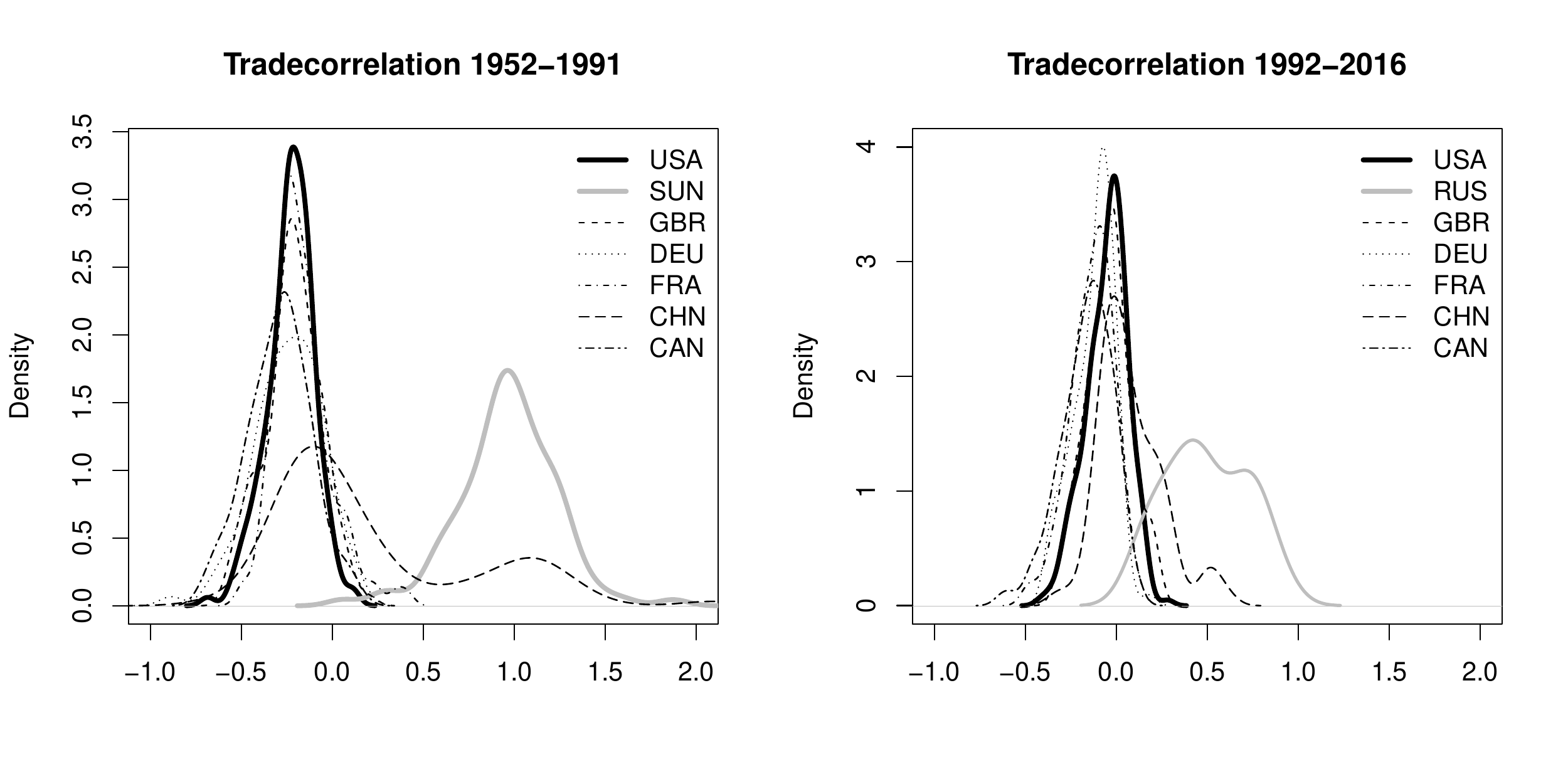}	
	\caption{Kernel Density Estimates of the Tradecorrelation Residuals for selected Countries in the Time Periods 1952-1991 (left) and 1992-2016 (right).}
	\label{densities}
\end{figure}
As shown in the previous section, the coefficient on the tradecorrelation is always positive and it is therefore natural to ask whether it is possible to identify trade clusters with higher or lower than expected trade flows. In order to do so, we attribute to each country those elements of the vector of estimated "tradecorrelation-residuals" ($\hat{\rho}\mathbf{W}\hat{\mathbf{u}}$) where the respective country is either the sender or the receiver. 
This is illustrated in Figure \ref{densities}. For the periods 1952-1991 (left) and 1992-2016 (right) kernel density estimates of the the estimated "tradecorrelation-residuals" for the most important countries are given. A strong difference in the distributions of the residuals can be found. Most notably the residuals are mostly negative for the United States (USA), Great Britain (GBR), Germany (DEU), France (FRA), and Canada (CAN) while the distribution for the Soviet Union (SUN) is shifted to the right and the one of China (CHN) is somewhat in-between In the period 1992-2016, the distributions are shifted closer together but the remarkable divide between the big western exporters and Russia (RUS) is still visible. This leads to the suggestion that the positive coefficient on tradecorrelation is partly driven by the differences of the tradeclusters of the western world and the one around the Soviet Union/Russia

\FloatBarrier

\FloatBarrier
\section{Conclusion}\label{conc}
In this paper we analysed the volume of international arms trade flows by employing a Network Disturbance Model. Seven different dependence structures are proposed and the best one is selected by the minimization of the AIC. Using this approach we find that a specification that relates the error term of an arms trade flow to the whole trade activity of the sending and receiving state works well in most years. This reveals that the network of international arms trade imposes a very complex dependence structure among the arms trade flows.

The analysis of the tradecorrelation coefficient shows that the correlation pattern among the flows rather stable and breaks only for the transition period after the end of the cold war. The development of the coefficients on the covariates shows how the influence of economic and political factors has changed with time.

\section*{Acknowledgements}
The authors gratefully acknowledge funding provided by the German Research Foundation (DFG) for the project \textit{International Trade of Arms: A Network Approach}.

\bibliographystyle{Chicago}
\bibliography{literature}

\begin{thebibliography}{}

\bibitem[\protect\citeauthoryear{Akerman and Seim}{Akerman and
  Seim}{2014}]{akerman2014}
Akerman, A. and A.~L. Seim (2014).
\newblock The global arms trade network 1950--2007.
\newblock {\em Journal of Comparative Economics\/}~{\em 42\/}(3), 535--551.

\bibitem[\protect\citeauthoryear{Ba}{Ba}{2017}]{ba2017}
Ba, H.-L.~K. (2017).
\newblock {\em Development, transition, and crisis in the international system,
  1870-2009}.
\newblock Ph.\ D. thesis, The University of North Carolina at Chapel Hill.

\bibitem[\protect\citeauthoryear{Bivand, Hauke, and Kossowski}{Bivand
  et~al.}{2013}]{bivand2013computing}
Bivand, R., J.~Hauke, and T.~Kossowski (2013).
\newblock Computing the jacobian in gaussian spatial autoregressive models: an
  illustrated comparison of available methods.
\newblock {\em Geographical Analysis\/}~{\em 45\/}(2), 150--179.

\bibitem[\protect\citeauthoryear{Bivand and Piras}{Bivand and
  Piras}{2015}]{Bivand2015}
Bivand, R. and G.~Piras (2015).
\newblock Comparing implementations of estimation methods for spatial
  econometrics.
\newblock {\em Journal of Statistical Software\/}~{\em 63\/}(18), 1--36.

\bibitem[\protect\citeauthoryear{Burnham and Anderson}{Burnham and
  Anderson}{2004}]{burnham2004}
Burnham, K.~P. and D.~R. Anderson (2004).
\newblock Multimodel inference: understanding aic and bic in model selection.
\newblock {\em Sociological Methods \& Research\/}~{\em 33\/}(2), 261--304.

\bibitem[\protect\citeauthoryear{Claeskens and Hjort}{Claeskens and
  Hjort}{2008}]{claeskens2008}
Claeskens, G. and N.~L. Hjort (2008).
\newblock {\em Model selection and model averaging}.
\newblock Cambridge: Cambridge University Press.

\bibitem[\protect\citeauthoryear{{Correlates of War Project}}{{Correlates of
  War Project}}{2017a}]{Defagr2016}
{Correlates of War Project} (2017a).
\newblock International military alliances, 1648-2012, version 4.1.
\newblock \url{http://www.correlatesofwar.org/data-sets/formal-alliances}.

\bibitem[\protect\citeauthoryear{{Correlates of War Project}}{{Correlates of
  War Project}}{2017b}]{cinc2017}
{Correlates of War Project} (2017b).
\newblock National material capabilities, 1816-2012, version 5.0.
\newblock
  \url{http://www.correlatesofwar.org/data-sets/national-material-capabilities}.

\bibitem[\protect\citeauthoryear{Csardi and Nepusz}{Csardi and
  Nepusz}{2006}]{csardi2006}
Csardi, G. and T.~Nepusz (2006).
\newblock The igraph software package for complex network research.
\newblock {\em InterJournal, Complex Systems\/}~{\em 1695\/}(5), 1--9.

\bibitem[\protect\citeauthoryear{Desmarais and Cranmer}{Desmarais and
  Cranmer}{2012}]{desmarais2012}
Desmarais, B.~A. and S.~J. Cranmer (2012).
\newblock Statistical inference for valued-edge networks: The generalized
  exponential random graph model.
\newblock {\em PloS one\/}~{\em 7\/}(1), e30136.

\bibitem[\protect\citeauthoryear{Frank and Strauss}{Frank and
  Strauss}{1986}]{frank1986}
Frank, O. and D.~Strauss (1986).
\newblock Markov graphs.
\newblock {\em Journal of the American Statistical Association\/}~{\em
  81\/}(395), 832--842.

\bibitem[\protect\citeauthoryear{Freeman}{Freeman}{2017}]{freeman2017}
Freeman, L.~C. (2017).
\newblock {\em Research methods in social network analysis}.
\newblock Routledge.

\bibitem[\protect\citeauthoryear{Gibler}{Gibler}{2009}]{gibler2008}
Gibler, D.~M. (2009).
\newblock {\em International military alliances, 1648-2008}.
\newblock Washington, DC: Congressional Quarterly Press.

\bibitem[\protect\citeauthoryear{Gleditsch}{Gleditsch}{2013a}]{gleditsch2013d}
Gleditsch, K.~S. (2013a).
\newblock Distance between capital cities.
\newblock \url{http://privatewww.essex.ac.uk/~ksg/data-5.html}.

\bibitem[\protect\citeauthoryear{Gleditsch}{Gleditsch}{2013b}]{gleditsch2013}
Gleditsch, K.~S. (2013b).
\newblock Expanded trade and gdp data.
\newblock \url{http://privatewww.essex.ac.uk/~ksg/exptradegdp.html}.

\bibitem[\protect\citeauthoryear{Handcock, Hunter, Butts, Goodreau, and
  Morris}{Handcock et~al.}{2008}]{handcock2008}
Handcock, M.~S., D.~R. Hunter, C.~T. Butts, S.~M. Goodreau, and M.~Morris
  (2008).
\newblock statnet: Software tools for the representation, visualization,
  analysis and simulation of network data.
\newblock {\em Journal of Statistical Software\/}~{\em 24\/}(1), 1548--7660.

\bibitem[\protect\citeauthoryear{Hays, Kachi, and Franzese}{Hays
  et~al.}{2010}]{hays2010}
Hays, J.~C., A.~Kachi, and R.~J. Franzese (2010).
\newblock A spatial model incorporating dynamic, endogenous network
  interdependence: A political science application.
\newblock {\em Statistical Methodology\/}~{\em 7\/}(3), 406--428.

\bibitem[\protect\citeauthoryear{Hlavac}{Hlavac}{2013}]{hlavac2013}
Hlavac, M. (2013).
\newblock stargazer: Latex code and ascii text for well-formatted regression
  and summary statistics tables.
\newblock \url{http://CRAN. R-project. org/package= stargazer}.

\bibitem[\protect\citeauthoryear{Holland and Leinhardt}{Holland and
  Leinhardt}{1981}]{holland1981}
Holland, P.~W. and S.~Leinhardt (1981).
\newblock An exponential family of probability distributions for directed
  graphs.
\newblock {\em Journal of the American Statistical Association\/}~{\em
  76\/}(373), 33--50.

\bibitem[\protect\citeauthoryear{Holtom, Bromley, and Simmel}{Holtom
  et~al.}{2012}]{holtom2012}
Holtom, P., M.~Bromley, and V.~Simmel (2012).
\newblock {\em Measuring international arms transfers}.
\newblock Stockholm International Peace Research Institute.

\bibitem[\protect\citeauthoryear{Jackson}{Jackson}{2010}]{jackson2010}
Jackson, M.~O. (2010).
\newblock {\em Social and economic networks}.
\newblock Princeton: Princeton University Press.

\bibitem[\protect\citeauthoryear{Kauermann, Haupt, and Kaufmann}{Kauermann
  et~al.}{2012}]{kauermann2012}
Kauermann, G., H.~Haupt, and N.~Kaufmann (2012).
\newblock A hitchhiker's view on spatial statistics and spatial econometrics
  for lattice data.
\newblock {\em Statistical Modelling\/}~{\em 12\/}(5), 419--440.

\bibitem[\protect\citeauthoryear{Kinne}{Kinne}{2016}]{Kinne2016}
Kinne, B.~J. (2016, may).
\newblock Agreeing to arm.
\newblock {\em Journal of Peace Research\/}~{\em 53\/}(3), 359--377.

\bibitem[\protect\citeauthoryear{Kolaczyk}{Kolaczyk}{2009}]{kolaczyk2009}
Kolaczyk, E.~D. (2009).
\newblock {\em Statistical analysis of network data. Methods and Models}.
\newblock New York: Springer-Verlag.

\bibitem[\protect\citeauthoryear{Krivitsky}{Krivitsky}{2012}]{krivitsky2012}
Krivitsky, P.~N. (2012).
\newblock Exponential-family random graph models for valued networks.
\newblock {\em Electronic Journal of Statistics\/}~{\em 6}, 1100--1128.

\bibitem[\protect\citeauthoryear{Leenders}{Leenders}{2002}]{leenders2002}
Leenders, R. T.~A. (2002).
\newblock Modeling social influence through network autocorrelation:
  constructing the weight matrix.
\newblock {\em Social networks\/}~{\em 24\/}(1), 21--47.

\bibitem[\protect\citeauthoryear{Leifeld, Cranmer, and Desmarais}{Leifeld
  et~al.}{2016}]{tnam}
Leifeld, P., S.~J. Cranmer, and B.~A. Desmarais (2016).
\newblock {\em xergm: Extensions for Exponential Random Graph Models}.
\newblock R package version 1.6.2.

\bibitem[\protect\citeauthoryear{LeSage and Pace}{LeSage and
  Pace}{2009}]{lesage2009}
LeSage, J.~P. and R.~K. Pace (2009).
\newblock {\em Introduction to spatial econometrics}.
\newblock London: Taylor and Francis.

\bibitem[\protect\citeauthoryear{Lusher, Koskinen, and Robins}{Lusher
  et~al.}{2012}]{lusher2012}
Lusher, D., J.~Koskinen, and G.~Robins (2012).
\newblock {\em Exponential random graph models for social networks: Theory,
  methods, and applications}.
\newblock Cambridge: Cambridge University Press.

\bibitem[\protect\citeauthoryear{Marshall}{Marshall}{2017}]{marshall2017}
Marshall, M.~G. (2017).
\newblock Polity iv project: Political regime characteristics and transitions,
  1800-2016.

\bibitem[\protect\citeauthoryear{Metz and Ingold}{Metz and
  Ingold}{2017}]{metz2017}
Metz, F. and K.~Ingold (2017).
\newblock Politics of the precautionary principle: assessing actors’
  preferences in water protection policy.
\newblock {\em Policy Sciences\/}~{\em 50\/}(4), 721--743.

\bibitem[\protect\citeauthoryear{Moran}{Moran}{1950}]{moran1950}
Moran, P.~A. (1950).
\newblock Notes on continuous stochastic phenomena.
\newblock {\em Biometrika\/}~{\em 37\/}(1/2), 17--23.

\bibitem[\protect\citeauthoryear{Moritz}{Moritz}{2016}]{Moritz2016}
Moritz, S. (2016).
\newblock {\em imputeTS: Time Series Missing Value Imputation}.
\newblock R package version 1.6.

\bibitem[\protect\citeauthoryear{Nelder and Mead}{Nelder and
  Mead}{1965}]{nelder1965simplex}
Nelder, J.~A. and R.~Mead (1965).
\newblock A simplex method for function minimization.
\newblock {\em The Computer Journal\/}~{\em 7\/}(4), 308--313.

\bibitem[\protect\citeauthoryear{{R Development Core Team}}{{R Development Core
  Team}}{2014}]{team2014r}
{R Development Core Team} (2014).
\newblock {\em R: A Language and Environment for Statistical Computing}.
\newblock Vienna, Austria: R Foundation for Statistical Computing.

\bibitem[\protect\citeauthoryear{Silk, Croft, Delahay, Hodgson, Weber, Boots,
  and McDonald}{Silk et~al.}{2017}]{silk2017}
Silk, M.~J., D.~P. Croft, R.~J. Delahay, D.~J. Hodgson, N.~Weber, M.~Boots, and
  R.~A. McDonald (2017).
\newblock The application of statistical network models in disease research.
\newblock {\em Methods in Ecology and Evolution\/}~{\em 8\/}(9), 1026--1041.

\bibitem[\protect\citeauthoryear{Singer, Bremer, and Stuckey}{Singer
  et~al.}{1972}]{singer1972}
Singer, J.~D., S.~Bremer, and J.~Stuckey (1972).
\newblock Capability distribution, uncertainty, and major power war, 1820-1965.
\newblock {\em Peace, War, and Numbers\/}~{\em 19}, 19--48.

\bibitem[\protect\citeauthoryear{{SIPRI}}{{SIPRI}}{2017a}]{sipridata2017}
{SIPRI} (2017a).
\newblock Arms transfers database.
\newblock \url{https://www.sipri.org/databases/armstransfers}.

\bibitem[\protect\citeauthoryear{{SIPRI}}{{SIPRI}}{2017b}]{siprimeth2017}
{SIPRI} (2017b).
\newblock Arms transfers database - methodology.
\newblock \url{https://www.sipri.org/databases/armstransfers/background}.

\bibitem[\protect\citeauthoryear{Thurner, Skyler, Kauermann, and
  Schmid}{Thurner et~al.}{2018}]{thurner2017}
Thurner, P.~W., C.~Skyler, G.~Kauermann, and C.~Schmid (2018).
\newblock The network of major conventional weapons transfers 1950-2013.
\newblock {\em Ms. LMU Munich, submitted\/}.

\bibitem[\protect\citeauthoryear{Wickham}{Wickham}{2016}]{wickham2016}
Wickham, H. (2016).
\newblock {\em ggplot2: elegant graphics for data analysis}.
\newblock New York: Springer-Verlag.

\bibitem[\protect\citeauthoryear{Willardson}{Willardson}{2013}]{willardson2013}
Willardson, S.~L. (2013).
\newblock {\em Under the influence Of arms: the foreign policy causes and
  consequences of arms transfers}.
\newblock Ph.\ D. thesis, University of Iowa.

\bibitem[\protect\citeauthoryear{{World Bank}}{{World Bank}}{2017}]{GDP2017}
{World Bank} (2017).
\newblock World bank open data, real gdp.
\newblock \url{http://data.worldbank.org/}.

\end{thebibliography}

\newpage
\pagenumbering{Roman}
\appendix
\FloatBarrier
\section{Annex}

\subsection{Descriptives}\label{descrannex}
Table \ref{coverage} gives the categories of arms that are included in the analysis. All types with explanations are taken from \citet{siprimeth2017}. The 171 countries that are included in our analysis can be found in Table \ref{count_inc}, together with the three-digit country codes that are used to abbreviate countries in the paper. In  addition to that, the time periods, for which we coded the countries as existent are included. Note that the SIPRI data set contains more than 171 arms trading entities but we excluded non-states and countries with no (reliable) covariates available. In the covariates some missings are present in the data. No time series of covariates for the selected countries is completely missing (those countries are excluded from the analysis) and the major share of them is complete but there are series with some missing values. This is  sometimes the case in the year $1990$ and/or $1991$ where the former socialist countries split up or had some transition time. In other cases values at the beginning or at the end of the series are missing. We have decided to impute the missing values via linear interpolation, using the \texttt{R} package \texttt{imputeTS} by \citet{Moritz2016}.
\begin{table}\footnotesize
	\centering
	\resizebox{\columnwidth}{!}{%
	\begin{tabular}{l p{11cm}}
		
		\hline
		Type & Explanation \\ 
		\hline
		Aircraft                                                                                                                                                   & All fixed-wing aircraft and helicopters, including unmanned aircraft  with a minimum loaded weight of 20 kg. Exceptions are microlight aircraft, powered and unpowered gliders and target drones.                                                                                                                                                                                                                                                                                                       \\ 
		Air-defence systems                                                                                                                                         & (a) All land-based surface-to-air missile systems, and (b) all anti-aircraft guns with a calibre of more than 40 mm or with multiple barrels with a combined caliber of at least 70 mm. This includes self-propelled systems on armoured or unarmoured chassis.                                                                                                                                                                                                                                            \\
		Anti-submarine warfare weapons                                                                                                                             & Rocket launchers, multiple rocket launchers and mortars for use against submarines, with a calibre equal to or above 100 mm.                                                                                                                                                                                                                                                                                                                                                                                      \\
		Armoured vehicles                                                                                                                                          & All vehicles with integral armour protection, including all types of tank, tank destroyer, armoured car, armoured personnel carrier, armoured support vehicle and infantry fighting vehicle. Vehicles with very light armour protection (such as trucks with an integral but lightly armoured cabin) are excluded.                                                                                                                                                                                               \\
		Artillery                                                                                                                                                  & Naval, fixed, self-propelled and towed guns, howitzers, multiple rocket launchers and mortars, with a calibre equal to or above 100 mm.                                                                                                                                                                                                                                                                                                                                                                        \\
		Engines                                                                                                                                                    & (a) Engines for military aircraft, for example, combat-capable aircraft, larger military transport and support aircraft, including large helicopters; (b) Engines for combat ships -,fast attack craft, corvettes, frigates, destroyers, cruisers, aircraft carriers and submarines; (c) Engines for most armoured vehicles - generally engines of more than 200 horsepower output.$^{*}$                                                                                                                            \\
		Missiles                                                                                                                                                   & (a) All powered, guided missiles and torpedoes, and (b) all unpowered but guided bombs and shells. This includes man-portable air defence systems  and portable guided anti-tank missiles. Unguided rockets, free-fall aerial munitions, anti-submarine rockets and target drones are excluded.                                                                                                                                                                                                         \\
		Sensors                                                                                                                                                    & (a) All land-, aircraft- and ship-based active (radar) and passive (e.g. electro-optical) surveillance systems with a range of at least 25 kilometres, with the exception of navigation and weather radars, (b) all fire-control radars, with the exception of range-only radars, and (c) anti-submarine warfare and anti-ship sonar systems for ships and helicopters.$^{*}$                                                                                                                                                                                                                                                                                                                                                                                                                                                                                                                                                                                 \\
		
		Satellites                                                                                                                                                   &  All reconnaissance and communications satellites.                                                                                                                                                                                                                                                                                                                                                                                                                                                                                                                                                                               \\
		Ships                                                                                                                                                      & (a) All ships with a standard tonnage of 100 tonnes or more, and (b) all ships armed with artillery of 100-mm calibre or more, torpedoes or guided missiles, and (c) all ships below 100 tonnes where the maximum speed (in kmh) multiplied with the full tonnage equals 3500 or more. Exceptions are most survey ships, tugs and some transport ships                                                                                                                                                           \\
		Other                                                                                                                                                      & (a) All turrets for armoured vehicles fitted with a gun of at least 12.7 mm calibre or with guided anti-tank missiles, (b) all turrets for ships fitted with a gun of at least 57-mm calibre, and (c) all turrets for ships fitted with multiple guns with a combined calibre of at least 57 mm, and (d) air refueling systems as used on tanker aircraft.$^{*}$                                                                                                                                                     \\ \hline
		\multicolumn{2}{p{17.5cm}}{$^{*}$In cases where the system is fitted on a platform (vehicle, aircraft or ship), the database only includes those systems that come from a different supplier from the supplier of the platform.}\\
		\multicolumn{2}{p{17.5cm}}{The Arms Transfers Database does not cover other military equipment such as small arms and light weapons (SALW) other than portable guided missiles such as man-portable air defence systems and guided anti-tank missiles. Trucks, artillery under 100-mm calibre, ammunition, support equipment and components (other than those mentioned above), repair and support services or technology transfers are also not included in the database.}
		\\
		\multicolumn{2}{p{15cm}}{Source: \citet{siprimeth2017}}
	\end{tabular}
}
	\caption{Types of Weapons included in the SIPRI Arms Trade Database}
	\label{coverage}
\end{table}
\FloatBarrier

\begin{table}[ht]
	\centering \footnotesize
	\resizebox{\columnwidth}{!}{%
	\begin{tabular}{lllllllll}
		\hline
		Country & Code & Included & Country & Code & Included &Country & Code & Included \\  
		\hline
		Afghanistan & AFG & 1950 - 2016 & German Dem. Rep. & GDR & 1950 - 1991 & Pakistan & PAK & 1950 - 2016 \\ 
		Albania & ALB & 1950 - 2016 & Germany & DEU & 1950 - 2016 & Panama & PAN & 1950 - 2016 \\ 
		Algeria & DZA & 1962 - 2016 & Ghana & GHA & 1957 - 2016 & Papua New Guin. & PNG & 1975 - 2016 \\ 
		Angola & AGO & 1975 - 2016 & Greece & GRC & 1950 - 2016 & Paraguay & PRY & 1950 - 2016 \\ 
		Argentina & ARG & 1950 - 2016 & Guatemala & GTM & 1950 - 2016 & Peru & PER & 1950 - 2016 \\ 
		Armenia & ARM & 1991 - 2016 & Guinea & GIN & 1958 - 2016 & Philippines & PHL & 1950 - 2016 \\ 
		Australia & AUS & 1950 - 2016 & Guinea-Bissau & GNB & 1973 - 2016 & Poland & POL & 1950 - 2016 \\ 
		Austria & AUT & 1950 - 2016 & Guyana & GUY & 1966 - 2016 & Portugal & PRT & 1950 - 2016 \\ 
		Azerbaijan & AZE & 1991 - 2016 & Haiti & HTI & 1950 - 2016 & Qatar & QAT & 1971 - 2016 \\ 
		Bahrain & BHR & 1971 - 2016 & Honduras & HND & 1950 - 2016 & Romania & ROM & 1950 - 2016 \\ 
		Bangladesh & BGD & 1971 - 2016 & Hungary & HUN & 1950 - 2016 & Russia & RUS & 1992 - 2016 \\ 
		Belarus & BLR & 1991 - 2016 & India & IND & 1950 - 2016 & Rwanda & RWA & 1962 - 2016 \\ 
		Belgium & BEL & 1950 - 2016 & Indonesia & IDN & 1950 - 2016 & Saudi Arabia & SAU & 1950 - 2016 \\ 
		Benin & BEN & 1961 - 2016 & Iran & IRN & 1950 - 2016 & Senegal & SEN & 1960 - 2016 \\ 
		Bhutan & BTN & 1950 - 2016 & Iraq & IRQ & 1950 - 2016 & Serbia & SRB & 1992 - 2016 \\ 
		Bolivia & BOL & 1950 - 2016 & Ireland & IRL & 1950 - 2016 & Sierra Leone & SLE & 1961 - 2016 \\ 
		Bosnia Herzegov. & BIH & 1992 - 2016 & Israel & ISR & 1950 - 2016 & Singapore & SGP & 1965 - 2016 \\ 
		Botswana & BWA & 1966 - 2016 & Italy & ITA & 1950 - 2016 & Slovakia & SVK & 1993 - 2016 \\ 
		Brazil & BRA & 1950 - 2016 & Jamaica & JAM & 1962 - 2016 & Slovenia & SVN & 1991 - 2016 \\ 
		Bulgaria & BGR & 1950 - 2016 & Japan & JPN & 1950 - 2016 & Solomon Islands & SLB & 1978 - 2016 \\ 
		Burkina Faso & BFA & 1960 - 2016 & Jordan & JOR & 1950 - 2016 & Somalia & SOM & 1960 - 2016 \\ 
		Burundi & BDI & 1962 - 2016 & Kazakhstan & KAZ & 1991 - 2016 & South Africa & ZAF & 1950 - 2016 \\ 
		Cambodia & KHM & 1953 - 2016 & Kenya & KEN & 1963 - 2016 & Soviet Union & SUN & 1950 - 1991 \\ 
		Cameroon & CMR & 1960 - 2016 & North Korea & PRK & 1950 - 2016 & Spain & ESP & 1950 - 2016 \\ 
		Canada & CAN & 1950 - 2016 & South Korea & KOR & 1950 - 2016 & Sri Lanka & LKA & 1950 - 2016 \\ 
		Cape Verde & CPV & 1975 - 2016 & Kuwait & KWT & 1961 - 2016 & Sudan & SDN & 1956 - 2016 \\ 
		Central Afr. Rep. & CAF & 1960 - 2016 & Kyrgyzstan & KGZ & 1991 - 2016 & Suriname & SUR & 1975 - 2016 \\ 
		Chad & TCD & 1960 - 2016 & Laos & LAO & 1950 - 2016 & Swaziland & SWZ & 1968 - 2016 \\ 
		Chile & CHL & 1950 - 2016 & Latvia & LVA & 1991 - 2016 & Sweden & SWE & 1950 - 2016 \\ 
		China & CHN & 1950 - 2016 & Lebanon & LBN & 1950 - 2016 & Switzerland & CHE & 1950 - 2016 \\ 
		Colombia & COL & 1950 - 2016 & Lesotho & LSO & 1966 - 2016 & Syria & SYR & 1950 - 2016 \\ 
		Comoros & COM & 1975 - 2016 & Liberia & LBR & 1950 - 2016 & Taiwan & TWN & 1950 - 2016 \\ 
		DR Congo & ZAR & 1960 - 2016 & Libya & LBY & 1951 - 2016 & Tajikistan & TJK & 1991 - 2016 \\ 
		Congo & COG & 1960 - 2016 & Lithuania & LTU & 1990 - 2016 & Tanzania & TZA & 1961 - 2016 \\ 
		Costa Rica & CRI & 1950 - 2016 & Luxembourg & LUX & 1950 - 2016 & Thailand & THA & 1950 - 2016 \\ 
		Cote dIvoire & CIV & 1960 - 2016 & Macedonia  & MKD & 1991 - 2016 & Timor-Leste & TMP & 2002 - 2016 \\ 
		Croatia & HRV & 1991 - 2016 & Madagascar & MDG & 1960 - 2016 & Togo & TGO & 1960 - 2016 \\ 
		Cuba & CUB & 1950 - 2016 & Malawi & MWI & 1964 - 2016 & Trinidad Tobago & TTO & 1962 - 2016 \\ 
		Cyprus & CYP & 1960 - 2016 & Malaysia & MYS & 1957 - 2016 & Tunisia & TUN & 1956 - 2016 \\ 
		Czech Republic & CZR & 1993 - 2016 & Mali & MLI & 1960 - 2016 & Turkey & TUR & 1950 - 2016 \\ 
		Czechoslovakia & CZE & 1950 - 1991 & Mauritania & MRT & 1960 - 2016 & Turkmenistan & TKM & 1991 - 2016 \\ 
		Denmark & DNK & 1950 - 2016 & Mauritius & MUS & 1968 - 2016 & Uganda & UGA & 1962 - 2016 \\ 
		Djibouti & DJI & 1977 - 2016 & Mexico & MEX & 1950 - 2016 & Ukraine & UKR & 1991 - 2016 \\ 
		Dominican Rep. & DOM & 1950 - 2016 & Moldova & MDA & 1991 - 2016 & Un. Arab Emirates & ARE & 1971 - 2016 \\ 
		Ecuador & ECU & 1950 - 2016 & Mongolia & MNG & 1950 - 2016 & United Kingdom & GBR & 1950 - 2016 \\ 
		Egypt & EGY & 1950 - 2016 & Morocco & MAR & 1956 - 2016 & United States & USA & 1950 - 2016 \\ 
		El Salvador & SLV & 1950 - 2016 & Mozambique & MOZ & 1975 - 2016 & Uruguay & URY & 1950 - 2016 \\ 
		Equatorial Guin. & GNQ & 1968 - 2016 & Myanmar & MMR & 1950 - 2016 & Uzbekistan & UZB & 1991 - 2016 \\ 
		Eritrea & ERI & 1993 - 2016 & Namibia & NAM & 1990 - 2016 & Venezuela & VEN & 1950 - 2016 \\ 
		Estonia & EST & 1991 - 2016 & Nepal & NPL & 1950 - 2016 & Vietnam & VNM & 1976 - 2016 \\ 
		Ethiopia & ETH & 1950 - 2016 & Netherlands & NLD & 1950 - 2016 & South Vietnam & SVM & 1950 - 1975 \\ 
		Fiji & FJI & 1970 - 2016 & New Zealand & NZL & 1950 - 2016 & Yemen & YEM & 1991 - 2016 \\ 
		Finland & FIN & 1950 - 2016 & Nicaragua & NIC & 1950 - 2016 & North Yemen & NYE & 1950 - 1991 \\ 
		France & FRA & 1950 - 2016 & Niger & NER & 1960 - 2016 & South Yemen & SYE & 1950 - 1991 \\ 
		Gabon & GAB & 1960 - 2016 & Nigeria & NGA & 1960 - 2016 & Yugoslavia & YUG & 1950 - 1992 \\ 
		Gambia & GMB & 1965 - 2016 & Norway & NOR & 1950 - 2016 & Zambia & ZMB & 1964 - 2016 \\ 
		Georgia & GEO & 1991 - 2016 & Oman & OMN & 1950 - 2016 & Zimbabwe & ZWE & 1950 - 2016 \\ 
		\hline
	\end{tabular}
}
	\caption{Countries included in the Analysis with three-digit Country Codes and Time-Period of Inclusion in the Model.}
	\label{count_inc}
\end{table}
\FloatBarrier

\subsection{Technical Details on the Estimation}\label{estimation_annex}
The estimation of the models is done in \texttt{R} (see \citealp{team2014r}). For models with one weight matrix, there is an implementation by \citet{bivand2013computing} (see also  \citealp{Bivand2015}), based on a two-step method. First, the profiled likelihood is maximized with respect to $\rho$, by using the simplex method of \citet{nelder1965simplex}. In the second step the other parameters are found by weighted least squares. Given those estimates, the concentrated likelihood is updated for $\rho$. The major computational burden comes with the calculation of the log-determinant of $(\mathbf{I}_n-\rho \mathbf{W})$. The problem is solved by using the eigenvalues of $\mathbf{W}$, i.e.\ $\log(|\mathbf{I}_n-\rho \mathbf{W}|)=\sum_{i=1}^{n}\log(1-\rho \lambda_i)$, with $\lambda_i$ being the eigenvalues of $\mathbf{W}$.

Further important packages used in this paper are the \texttt{igraph} package by \citet{csardi2006} as well as the \texttt{statnet} suite by \citet{handcock2008} for the visualization of networks. For the Figures and Tables the packages \texttt{ggplot2} by \citet{wickham2016} and \texttt{stargazer} by \citet{hlavac2013} are used.

\FloatBarrier
\subsection{Optimal Cutoff for the Spatial Distance}\label{cutoffrannex}
\begin{figure}[t!]
	\centering
	
	\includegraphics[trim={0cm 0cm 0cm 0cm},clip,scale=0.5]{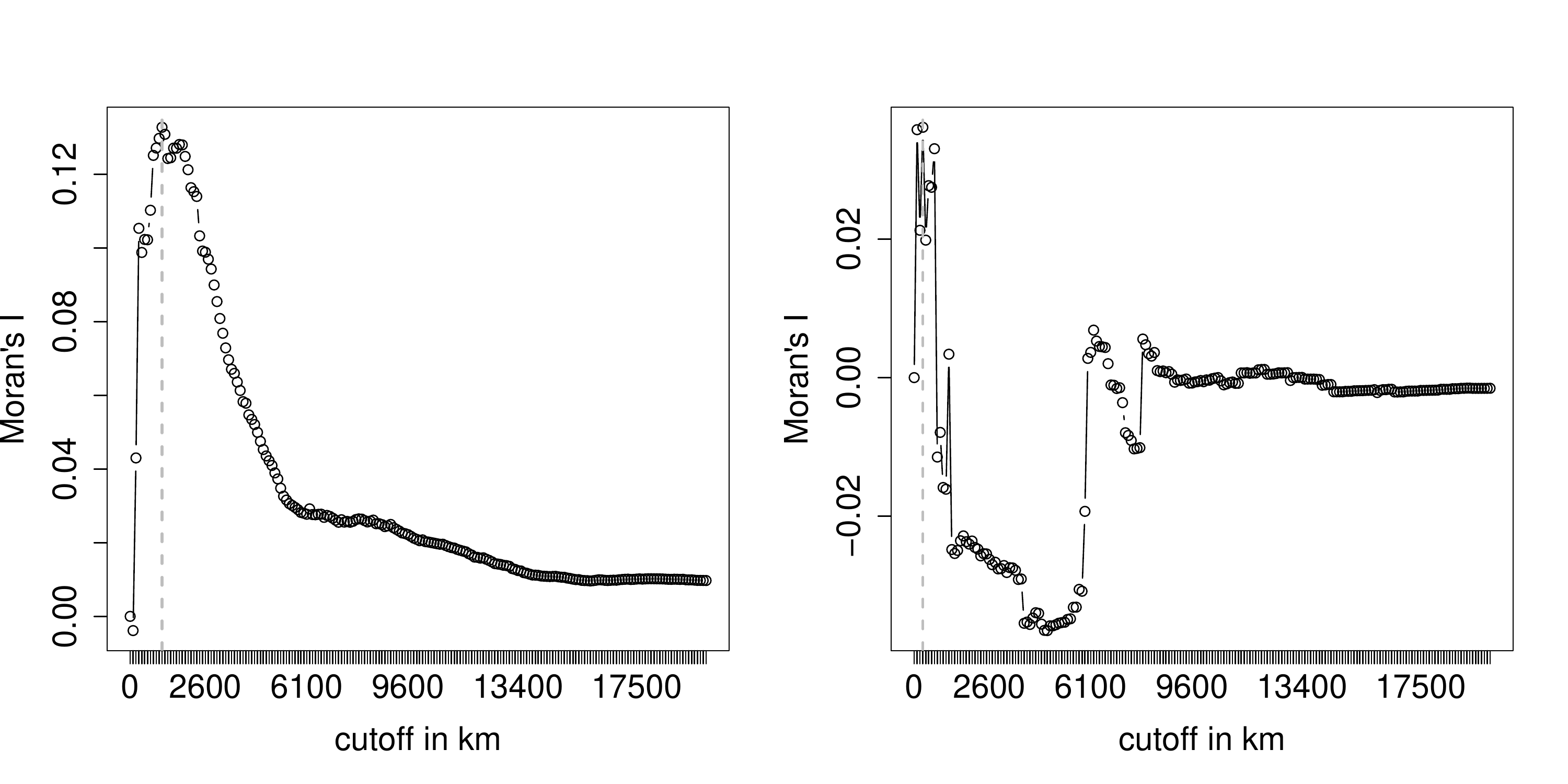}

	\caption{Moran's $I$ for different Threshold Levels $c$ on a Grid. Receiver related Tradecorrelation on the left and Sender related Tradecorrelation on the right.}
	\label{cutoff}
\end{figure}

The optimal cutoff distances are found by maximization of Moran's $I$ (see \citealp{moran1950}) for the whole time period. In order to do so we construct a block-diagonal weight matrix for all observations and time periods and calculate Moran's $I$ for varying threshold levels. The distance is discretized on a grid, ranging from $0$ km to $20\,000$ km, incrementing in steps of $100$ km. The highest value for Moran's $I$ is found for  $c=1\,100$ km for the weight matrix that relates the distances of the receiver ($\mathcal{N}_{5,import}$) and $c=300$ km for the weight matrix that relates the distances of the sender ($\mathcal{N}_{5,export}$). Note that we have chosen the maximum value in the sender-related model, although there is also a minimum value that is comparable in absolute numbers and could be interpreted as negative spatial correlation. This value is however realized for a cutoff value around $4\,600$ km which seems to be unrealistic (e.g.\ the distance between Moscow and London is roughly $2\,500$ km). The result is visualized in Figure \ref{cutoff}.

\FloatBarrier

\end{document}